\documentclass[]{jfm}

\usepackage{graphicx}
\usepackage{newtxtext}
\usepackage{newtxmath}
\usepackage{natbib}
\usepackage{hyperref}
\hypersetup{
    colorlinks=true,
    citecolor=blue,
    linkcolor=blue,
    anchorcolor=green,
    urlcolor=magenta,
    breaklinks=true
}
\usepackage{tikz}
\usetikzlibrary{quantikz2}
\usepackage{bm}
\newtheorem{definition}{Definition}
\newtheorem{theorem}{Theorem}

\title{Quantum implicit representation of vortex filaments in turbulence}

\author{Chenjia Zhu\aff{1},
  Ziteng Wang\aff{1}
  \corresp{Co-first author},
  Shiying Xiong\aff{1}
    \corresp{\email{shiying.xiong@zju.edu.cn}},
  Yaomin Zhao\aff{2,3}
 \and Yue Yang\aff{2,3}}

\affiliation{\aff{1}Department of Engineering Mechanics, School of Aeronautics and Astronautics, Zhejiang University, Hangzhou, 310027, China
\aff{2}State Key Laboratory for Turbulence and Complex Systems, College of Engineering, Peking University, Beijing, 100871, China
\aff{3}HEDPS-CAPT, Peking University, Beijing, 100871, China}

\begin{document}

\maketitle
\begin{abstract}
Entangled vortex filaments are essential to turbulence, serving as coherent structures that govern nonlinear fluid dynamics and support the reconstruction of fluid fields to reveal statistical properties. This study introduces an quantum implicit representation of vortex filaments in turbulence, employing a level-set method that models the filaments as the intersection of the real and imaginary zero iso-surfaces of a complex scalar field. Describing the fluid field via the scalar field offers distinct advantages in capturing complex structures, topological properties, and fluid dynamics, while opening new avenues for innovative solutions through quantum computing platforms. The representation is reformulated into an eigenvalue problem for Hermitian matrices, enabling the conversion of velocity fields into complex scalar fields that embed the vortex filaments. The resulting optimization is addressed using a variational quantum eigensolver, with Pauli operator truncation and deep learning techniques applied to improve efficiency and reduce noise. The proposed quantum framework achieves a near-linear time complexity and a exponential storage reduction while maintaining a balance of accuracy, robustness, and versatility, presenting a promising tool for turbulence analysis, vortex dynamics research, and machine learning dataset generation.
% The proposed quantum framework achieves speedups while maintaining a balance of accuracy, robustness, and versatility, presenting a promising tool for turbulence analysis, vortex dynamics research, machine learning dataset generation, and compact experimental data storage.
\end{abstract}

\begin{keywords}
vortex dynamics; turbulence modelling; topological fluid dynamics
% 等下要删掉的
% Authors should not enter keywords on the manuscript, as these must be chosen by the author during the online submission process and will then be added during the typesetting process (see \href{https://www.cambridge.org/core/journals/journal-of-fluid-mechanics/information/list-of-keywords}{Keyword PDF} for the full list).  Other classifications will be added at the same time.
\end{keywords}
% {\bf MSC Codes }  {\it(Optional)} Please enter your MSC Codes here

\section{Introduction}
\label{sec:introduction}
Turbulent and transitional flows are inherently chaotic across both spatial and temporal scales, yet they exhibit significant coherence through the presence of organized structures~\citep{She1990intermittent}. These coherent structures, particularly in regions of high vorticity, often manifest as tube-like formations, with velocity fields spiraling around them, as evidenced by both experimental \citep{Kuo1972experiment} and numerical investigations \citep{Siggia1981numerical}. Vortex tubes are integral to turbulence dynamics \citep{Jimenez1993structure, Jimenez1998characteristics}, with their evolution and interactions driving processes such as flux transport \citep{Pullin2000Mixing}, mixing \citep{Hussain1986Structure, Brown1974Density}, and the energy cascade \citep{Yao2020Cascade}.

Vortex filaments, conceptualized as infinitesimally thin vortex tubes, serve as an effective tool for characterizing turbulence, as they represent the coherent structures that govern the complex nonlinear dynamics of turbulent flows \citep{shen2024designing}. The vortex filament method, distinguished by its capability to control filament morphology, spatial distribution, and thickness, provides a versatile framework for capturing the multiscale nature of turbulence \citep{Pullin1998vortex}.
In turbulent flows, the motion of vortex filaments is driven by induced velocities, as described by the Biot--Savart (BS) law \citep{Davidson2004}, which links the vorticity distribution along the filament to the resulting velocity field at any point. This relationship forms the basis for advanced models that describe vortex dynamics, including stretching, twisting, and reconnection.

However, the vortex filament method encounters challenges when dealing with high filament densities and complex topological changes. The accurate simulation of vortex filament reconnection and interaction is a complex task, even with the use of regularization techniques like the Rosenhead--Moore (RM) model \citep{saffman1995vortex}. Another major challenge in the vortex filament method is the extraction of representative filaments from flow fields \citep{Gunther2018state}. One common approach involves constructing filaments based on velocity field and other physical properties, but this often leads to combinatorial optimization problems~\citep{lensgraf2017improved}.

To address these limitations, \citet{Weibmann2014} introduces an implicit representation of vortex filaments using the zero level set of a complex function, reformulating the problem as a continuous functional optimization. This approach offers several advantages, including the efficient representation of complex structures and flexibility in capturing topological dynamics~\citep{Yang2021clebsch,Xiong2022Clebsch}. Scalar field-based methods for representing lower-dimensional structures have been widely adopted in other areas of fluid dynamics, such as two-phase flow simulations in compressible \citep{Mulder1992computing} and incompressible \citep{Sussman1994level} fluids, and implicit vortex simulation \citep{Ishida2022hidden}. Despite its flexibility, when it comes to high-dimensional eigenvalue problems, the method can become processing-heavy, particularly for large-scale flow data. 

Quantum computing, however, presents a promising alternative. Leveraging the principles of superposition, entanglement, and interference~\citep{Nielsen2010quantum,Horowitz2019quantum}, it has the potential to offer computational advantages over classical methods, particularly for problems that involve complex, nonlinear dynamics~\citep{Liao2022threebody,Li2018clean}, and has drawn attention for accelerating computational tasks across scientific domains. 
In the context of fluid dynamics, quantum computing is expected to offer efficiency improvements over classical computational fluid dynamics (CFD) methods in specific scenarios. Several quantum algorithms have already been developed for simulating fluid dynamics \citep{Bharadwaj2024Simulating}, including studies on lattice Boltzmann equation \citep{Itani2024quantum}, Navier--Stokes (NS) equation with quantum linear systems algorithms (QLSA) \citep{Bharadwaj2023hybrid} and variational quantum algorithm (VQA) \citep{Ingelmann2024two}, and scalar transportation equation \citep{Lu2024quantum}. However, these algorithms have yet to provide a universally applicable framework for tackling the complex, nonlinear interactions that define real-world turbulent flows due to the linear nature of quantum mechanics in certain aspects \citep{Liu2021Efficient, Giannakis2022embedding}. In contrast, our current work focuses on the specific fluid dynamics challenge of vortex filament extraction, a problem particularly well-suited to the strengths of quantum computing.

We investigate the application of quantum algorithms to the extraction of vortex filaments, extending the implicit approach. %To make this problem amenable to quantum computation, we define an Ansatz that can be implemented as a sequence of quantum gates on a quantum device. The quantum state is prepared through a parameterized unitary transformation, represented by a general parameterized unitary operator acting on the ground-state qubits. 我觉得这一段和VQE有关，其他方法关系并不是很大，放后面了。
In our framework, the classical optimization problem is reformulated as a ground-state search problem, which corresponds to finding the minimum eigenvalue of a Hermitian matrix. Such problems can be addressed using variational methods like the variational quantum algorithm (VQA) \citep{Cerezo2021variational} and the variational quantum eigensolver (VQE) \citep{Tilly2022variational}, as well as exact approaches such as adiabatic quantum computation (AQC) \citep{albash2018adiabatic} and the quantum Lanczos method \citep{kirby2023exact}. While AQC and the quantum Lanczos method can obtain the exact ground state, they come with practical challenges; AQC requires stringent physical conditions, whereas the quantum Lanczos method depends on a well-chosen initial state. We adopt VQE as the primary quantum algorithm in our study given its straightforward implementation, adaptable ansatz for target systems, and inherent noise resilience. In our approach, the ground state is approximated using a parameterized quantum circuit, where the parameters are optimized by a classical optimizer to minimize the energy expectation value.
% 这个讨论是放在introduction里面好还是放在3.2更好？

The method proceeds as follows. First, we formulate an optimization problem that links the circulation of the velocity field to the winding number of a complex scalar field, enabling the reconstruction of the scalar field from the velocity data. This problem is discretized on a staggered grid and posed as an eigenvalue problem of a Hermitian matrix. A VQE is then applied to solve the minimum eigenvalue and eigenvector, which encodes the velocity field in a quantum state. Finally, the quantum state is measured, and a neural network is used to extract vortex filaments, providing a reduced-dimensional representation of the velocity field.
We validate the method by conducting numerical experiments across a range of flow scenarios, including random vortex distributions, leapfrogging vortex structures, turbulence, and knotted vortex tubes. These experiments assess the accuracy, efficiency, and robustness of the method in the presence of noise and perturbations.

% 我们的quantum implicit representation 是我们所知首个将量子计算应用于湍流分析的方法，将量子计算扩展到三维湍流流场的分析中。得益于量子计算，我们的方法此相较于传统的高斯消元法（$O(n^3)$）和Lanczos算法（$O(mn^2)$）在特征向量求解时实现了接近线性复杂度，并利用量子态编码指数级减少存储需求。最后，为了应对量子计算中的噪声问题，我们还结合了机器学习技术过滤量子计算结果中的噪声和误差，显著提高了涡丝提取的准确性和鲁棒性，大大扩展了量子计算在流体动力学中的适用范围。
To the best of our knowledge, our quantum implicit representation is the first methodology to successfully apply quantum computing to turbulence analysis, thereby extending quantum computational capabilities to fully three-dimensional turbulent flow fields. 
%By leveraging quantum algorithms, our method offering a near-linear complexity and an exponential data compression allowing for the efficient solution of this optimization problem. Moreover, our method exponentially reduces storage requirements by exploiting the relationship between qubits and classical bits. To address the challenges posed by quantum noise, we also integrate machine learning techniques to filter both noise and errors in quantum outputs, thereby significantly enhancing the accuracy and robustness of vortex filament extraction.
% Our quantum implicit representation demonstrates advantages throughout the entire procedure. By representing vortex filaments as a continuous complex scalar field, we bypass the combinatorial optimization challenges inherent in traditional approaches. The quantum algorithm achieves potential exponential speedup through the parallel processing capabilities inherent in quantum principles, while offering a low-dimensional storage method for fluid dynamics data by leveraging the ability of qubits to store information in superpositions. Once the quantum state is obtained, deep learning algorithms can be employed to identify zero-level isosurfaces for vortex filament extraction, improving noise reduction and extraction efficiency.
Beyond facilitating vortex filament extraction, this approach also lays a solid foundation for studying vortex dynamics and interactions. Precise vortex filament positions provide reliable initial conditions for vortex-based fluid simulations \citep{Brochu2012linear-time, Golas2012large-scale}. The extracted vortex filaments and scalar fields themselves serve as high-quality datasets for machine learning proposes, which often directly learns vortex structures and behaviors. By providing physically meaningful and well-structured data, our method enhances the accuracy of simulations and predictions in applications such as leading-edge vortex formation \citep{Lee2022leading} and vortex wake dynamics \citep{Ribeiro2023Machine}. The extracted vortices can also aid in optimizing drag reduction \citep{Xia2024Drag}, vibration control \citep{Bai2022machine}, and other critical aspects of aerospace engineering \citep{Mitchell2001research}. Furthermore, our method exponentially reduces storage requirements by exploiting the fact that $n$ qubits can encode information equivalent to $2^n$ classical bits, enabling efficient storage of experimental data such as time-resolved particle image velocimetry (PIV) measurements.

The structure of the paper is as follows: In \S \ref{sec:Implicit}, we introduce the implicit representation of vortex filaments in turbulence, using a complex scalar field to characterize vorticity through its winding number. This approach provides a flexible framework for modeling vortex dynamics and interactions. In \S \ref{sec:overview}, we present the quantum approach to implicit vortex representation, detailing the discretization of the optimization problem, the quantum algorithm used to solve it, and the extraction of vortex filaments. In \S \ref{sec:implement}, we discuss the implementation details of the quantum algorithms, including complexity analysis and parameter selection. In \S \ref{sec:results}, we present the results of applying the method to various flow configurations. Finally, in \S \ref{sec:conclusion}, we summarize the advantages and limitations of the approach and provide recommendations for future research.

\section{Implicit representation of vortex filaments in turbulence}
\label{sec:Implicit}
\subsection{Vortex filament representation} 
The vortex filament representation characterizes vorticity as discrete line singularities with circulation confined to the filament axis, effectively capturing the fundamental features of turbulent flows, including vortex generation, interaction, and dissipation. This approach provides a computationally efficient framework for analyzing vortical dynamics and energy transfer, particularly in high Reynolds number regimes, where turbulence is governed by the intricate interactions of vortex filaments.

In this representation, vorticity \( \bm{\omega} = \bm{\nabla} \times \bm{u} \), where \( \bm{u} \) is the velocity field, is concentrated along one-dimensional curves, known as vortex filaments. The physical domain is an open region \( \Omega \subset \mathbb{R}^3 \), and the set of \( m \) closed, time-dependent vortex filaments is described by~\citep[see][]{Ishida2022hidden}
\begin{equation}
\bm{\gamma}(:, t) : \bigsqcup_{i=1}^m S^1 \to \Omega,
\end{equation}
where \( \bigsqcup \) denotes a disjoint union, \( S^1 \) represents the topological circle, and \( \bm{\gamma} \) maps each vortex filament into the domain \( \Omega \). This framework provides a means to model the essential dynamics of turbulence. Specifically, ~\citet{shen2024designing} uses quantum vortex tangles to generate synthetic turbulence by combining vortex filaments, effectively capturing the key characteristics of turbulent flows. The ability to vary vortex filament thickness further facilitates the investigation of energy cascades and extreme events within turbulence dynamics.

Accurate representation of vortex structures is crucial for understanding vortex morphology~\citep{Xiong2019}. Vortex filaments, as the limiting case of vortex surfaces~\citep{Xiong2017}, form a network of elongated, intertwined structures. In contrast, iso-surfaces of the vorticity magnitude \( |\bm{\omega}| \) typically exhibit shorter, tube-like shapes, often referred to as ``vortex worms''~\citep{Jimenez1993structure}. This distinction becomes particularly pronounced during the energy cascade, where \( |\bm{\omega}| \) iso-surfaces fragment into smaller structures. However, such fragmentation provides only a partial view of vortex dynamics. Key processes in the energy cascade, including stretching, twisting, and reconnection, are more accurately represented by the continuous deformation of vortex filaments.

The motion of vortex filaments is governed by the kinematic equation
\begin{equation}
\frac{\partial \bm \gamma}{\partial t}(s, t) = \bm{u}_\gamma(s, t), \quad s \in \bigsqcup S^1, \; t \in \mathbb{R},
\end{equation}
where \( \bm{u}_\gamma(s, t) \) is the velocity of the filament. The BS law describes the self-induced motion of the vortex
\begin{equation}
\bm{u}^\text{BS}_\gamma(s) = \frac{\Gamma}{4\pi} \int_\gamma \frac{\bm \gamma'(s') \times [\bm \gamma(s) - \bm \gamma(s')]}{|\bm \gamma(s) - \bm \gamma(s')|^3} \, \mathrm{d}s',
\end{equation}
where \( \Gamma \) is the vortex strength, and \( \mathrm{d}s' \) is the arclength element. To regularize the singularity in the BS law, the RM model \citep{saffman1995vortex} introduces a core size \( a \), modifying the velocity to
\begin{equation}
\bm{u}^\text{RM}_\gamma(s) = \frac{\Gamma}{4\pi} \int_\gamma \frac{\bm \gamma'(s') \times [\bm \gamma(s) - \bm \gamma(s')]}{(\sqrt{[e^{-3/2}a^2 + |\bm \gamma(s) - \bm \gamma(s')|^2}]^3} \, \mathrm{d}s',
\label{eq:RM}
\end{equation}
which reduces to the BS law in the limit as \( a \to 0 \). These equations describe vortex transport and interactions, providing a link between local filament dynamics and the large-scale behavior of turbulence.

Despite its advantages, the vortex filament representation faces challenges in handling high filament densities and complex topological changes. These issues arise from the simplicity of the mathematical formulation and the limitations of existing numerical techniques. For example, reconnection algorithms integral to capturing filament dynamics often rely on subjective criteria, lacking the precision and rigor needed for high-fidelity simulations.   

\subsection{Implicit vortex filament representation}
%这边是不是能提一下用隐式方法更适合量子计算？比如说在这里就提及这个这个问题是Hermite的，或者说隐式方法它的编码长度不会变，只和网格有关，因此不需要在计算的过程中对量子电路进行调整；而显式表征会涉及连接等过程，不仅处理困难，而且需要对量子电路的规模进行不断调整，这对于量子计算是不利的。

\begin{figure}
    \centering
    \includegraphics[width=\linewidth]{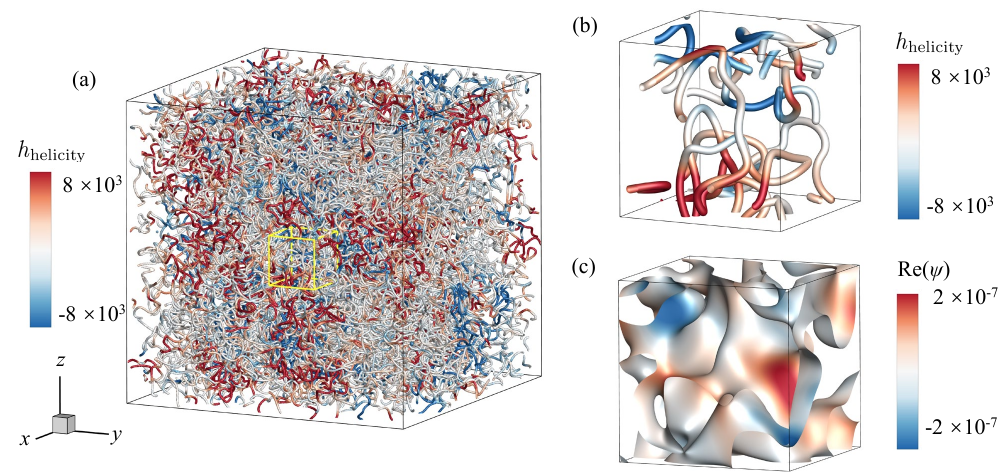}
    \caption{Vortex filaments and the complex scalar field $\psi$ in a three-dimensional turbulent flow. $h_{\textrm{helicity}}$ denotes the helicity density.} (a) Vortex filaments extracted from a three-dimensional turbulent flow, colored by helicity. (b) (c) Detailed views of the filaments and iso-surfaces of $\mathrm{Im}(\psi) = 0$ within the highlighted region (bright yellow box) in (a), colored by helicity and $\mathrm{Re}(\psi)$, respectively. The detailed views illustrate the complex entanglement of vortices in turbulent flows. 
    \label{fig:3DTurbA}
\end{figure}

% 隐式表征通过使用levelset可以提供更好的灵活性和适配性。一方面其通过使用levelset解决显式表征的局限性，解决对分裂和重连的描述问题；另一方面，隐式表征更适应我们的量子算法。隐式表征的编码数量永远和网格数量相等，不用在计算过程中调整量子电路的规模；同时，根据我们后面的推导，使用隐式表征能得到Hermite矩阵的形式，天然适合量子算法
In explicit methods, vortex filaments are represented by discrete points or lines, requiring continuous updates at each time step as their position and properties are recalculated. In contrast, implicit vortex filaments are defined as zero-isosurfaces of a scalar field, with their evolution inherently linked to changes in the scalar field itself. As the scalar field evolves, the zero-isosurface representing the filament adjusts automatically, eliminating the need for explicit updates of each individual point. The computational complexity inherent in explicit methods naturally suggests the potential of quantum computing, particularly given that a complex scalar field can be interpreted as a quantum state, aligning seamlessly with the foundational principles of quantum computation.

Before delving into our proposed quantum implicit vortex filament theoretical framework, we present a representative case to demonstrate its application in capturing the complex dynamics of chaotic turbulence. Figure~\ref{fig:3DTurbA} shows vortex filament extraction from a three-dimensional turbulent flow, where the input is a randomly generated velocity field. Figure~\ref{fig:3DTurbA}(a) illustrates the vortex filaments, colored by helicity, with a yellow box highlighting a region of interest for further analysis. Figure~\ref{fig:3DTurbA}(b) zooms in on this highlighted region, providing a detailed view of the filaments colored by helicity, capturing the intricate entanglement of the vortices. Figure~\ref{fig:3DTurbA}(c) focuses on the iso-surfaces of \( \text{Im}(\psi) = 0 \) within the same region, with the filaments colored by the real part of \( \psi \), offering a further visualization of the complex interactions between vortices in the flow.

Let \( \psi \) be a complex scalar field in \( \Omega \), with \( \partial \Sigma \) being a closed curve. The winding number of \(\psi\) around \(\partial \Sigma\) is a topological invariant that quantifies the total phase rotation of \(\psi\) along the curve. It is defined as  
\begin{equation}
n_{w} = \frac{1}{2\pi} \int_{\partial \Sigma} \bm{\nabla} (\arg \psi) \cdot \mathrm{d}\bm{l},
\end{equation}
where \(\arg \psi\) represents the phase of the complex scalar field, such that \( \psi = |\psi| e^{\mathrm{i}\theta} \) and \( \theta = \arg \psi \), and \(\bm{\nabla} (\arg \psi)\) is the phase gradient, capturing its spatial variation along the curve. The term \(\mathrm{d}\bm{l}\) is an infinitesimal vector element along the curve \(\partial \Sigma\), specifying the direction of integration. This integral measures the total phase change normalized by \( 2\pi \), counting the number of complete rotations around the curve. Since the phase of \( \psi \) returns to its original value after one full loop unless there is a singularity inside the curve, a non-zero winding number only occurs where the zero-level set of \( \psi \) exists, indicating that phase rotation does not return to zero. A non-zero winding number thus indicates the presence of one or more vortex filaments enclosed by \( \partial \Sigma \), with its magnitude reflecting the total circulation strength.

Alternatively, the winding number can also be expressed in a mathematically equivalent form that directly incorporates the complex scalar field \( \psi \) and its gradient, given by
\begin{equation}
n_{w} = \frac{1}{2\pi} \int_{\partial \Sigma} \frac{\langle \bm \nabla \psi, \mathrm{i} \psi \rangle}{|\psi|^2} \cdot \mathrm{d}\bm l,
\label{eq:winding}
\end{equation}
where \(\langle \bm{\nabla} \psi, \mathrm{i} \psi \rangle\) represents the inner product (or projection) of the gradient of \(\psi\) onto the imaginary axis, scaled by \(\psi\) itself. This isolates the change in the phase of \(\psi\), discarding changes in magnitude. Dividing by \(|\psi|^2\) normalizes \(\psi\), ensuring only the phase rotation contributes.

The winding number, as a topological invariant, provides a useful tool for characterizing vortex fields in wave-function-based representations. It serves as an analogue to the circulation of a velocity field in classical fluid mechanics, establishing a connection between topological properties and dynamical behavior. For a closed curve \( \partial \Sigma \), the circulation of the velocity field \( \bm{u} \) is defined as
\begin{equation}
\Gamma_u = \int_{\partial \Sigma} \bm u \cdot \mathrm{d}\bm l,
\label{eq:vortVel}
\end{equation}
where \( \bm{u} \) represents the velocity field. By Stokes' theorem, this circulation corresponds to the flux of vorticity through any surface \( \Sigma \) that is bounded by the curve \( \partial \Sigma \).

From equations \eqref{eq:winding} and \eqref{eq:vortVel}, the winding number of the complex scalar field and the circulation of the velocity field are both defined for any closed curve, thereby linking these two quantities. Following the approach of \citet{Weibmann2014}, we approximate the winding number and circulation as proportional, with the relation
\begin{equation}
\Gamma_{u} \approx h n_{w},
\label{eq:ghn}
\end{equation}  
where \( h \) is a proportionality constant that connects the two.

Equation \eqref{eq:ghn} is not directly suitable for numerical computation, so we reformulate it as an optimization problem to facilitate implementation. We substitute the expressions for \( n_w \) and \( \Gamma_u \) from equations \eqref{eq:winding} and \eqref{eq:vortVel} into \eqref{eq:ghn}, respectively. Then, we rearrange the terms, moving all non-zero terms to the left-hand side, yielding the following equation:
\begin{equation}
\frac{1}{2\pi} \int_{\p \Sigma} \frac{\langle \bm \nabla \psi - \mathrm{i} \hat{\bm u} \psi, \mathrm{i} \psi \rangle}{|\psi|^2} \cdot \mathrm{d}\bm l \approx 0,
\label{eq:def2}
\end{equation}
where \( \hat{\bm u} = 2\pi \bm u / h \). To make this equation hold, we pose the following optimization problem
\begin{equation}
\min_{\psi} \frac{1}{2} \|\bm \nabla \psi - \mathrm{i} \hat{\bm u} \psi\|^2, ~~\text{s.t.} ~~ \|\psi\|^2 = 1.
\label{eq:optpsi}
\end{equation}
Equation \eqref{eq:optpsi} represents a linear functional problem with a normalization constraint, which can be solved numerically on a discrete grid.

\begin{figure}
    \centering
    \includegraphics[width=1.0\textwidth]{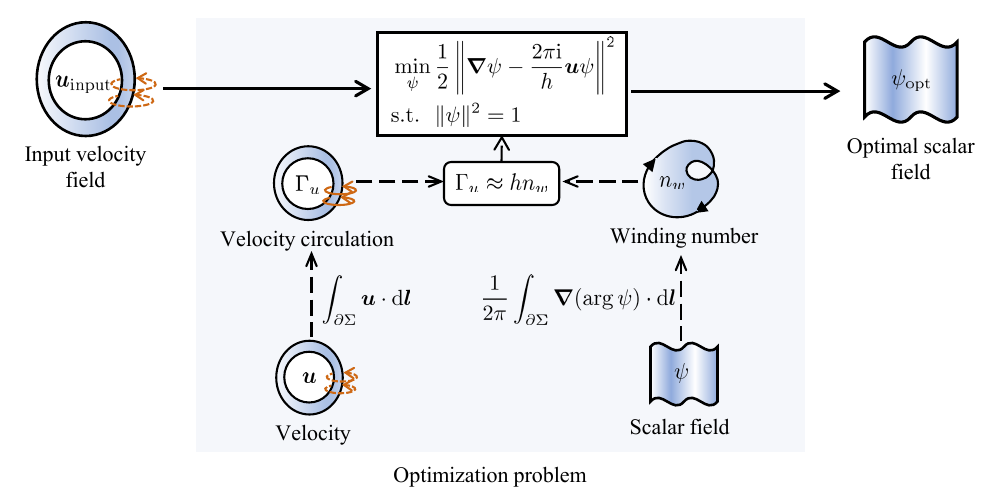}
    \caption{Schematic of the process for computing the complex scalar field from a given velocity field. The optimization problem is formulated using an arbitrary closed curve in $\Omega$, where the winding number and circulation are related through the expression in equation \eqref{eq:ghn}. The resulting complex scalar field approximates the velocity field, with its zero-level set corresponding to vortex filaments of strength \(h\).}
    \label{fig:optmization}
\end{figure}

Figure~\ref{fig:optmization} illustrates the procedure for computing the complex scalar field from a given velocity field. The process begins with a closed curve in \(\Omega\), where the winding number, defined in \eqref{eq:winding}, is derived from the complex scalar field, and the circulation, defined in \eqref{eq:vortVel}, is computed from the velocity field. Although these quantities are related conceptually, they are not typically directly proportional. To address this, the problem is reformulated as an optimization problem, as described in \eqref{eq:optpsi}, which aims to align the winding number and circulation as closely as possible. By solving this optimization problem for a given velocity field, a complex scalar function is obtained, which serves as an approximate representation of the velocity field. The zero-level set of this scalar field is interpreted as vortex filaments with strength \(h\), and the reconstructed vorticity flux across any surface is approximated to match the vorticity flux of the original velocity field.

Once the complex scalar field is constructed, we define the implicit representation of vortex filaments as follows.
\begin{definition}[Implicit representation of vortex filaments]  
\label{defn:implicit}
Given a complex scalar field \( \psi \), the implicit representation of vortex filaments is defined such that the vorticity is represented as a singular distribution at the zero-level set of \( \psi \), i.e., where \( \operatorname{Re}(\psi) = \operatorname{Imag}(\psi) = 0 \). The strength of the vortex filament is then given by \( h n_w \), where \( n_w \) is the winding number of \( \psi \) at that point, as defined by \eqref{eq:winding}.
\end{definition}

\begin{figure}
    \centering
    \includegraphics[width=1.\textwidth]{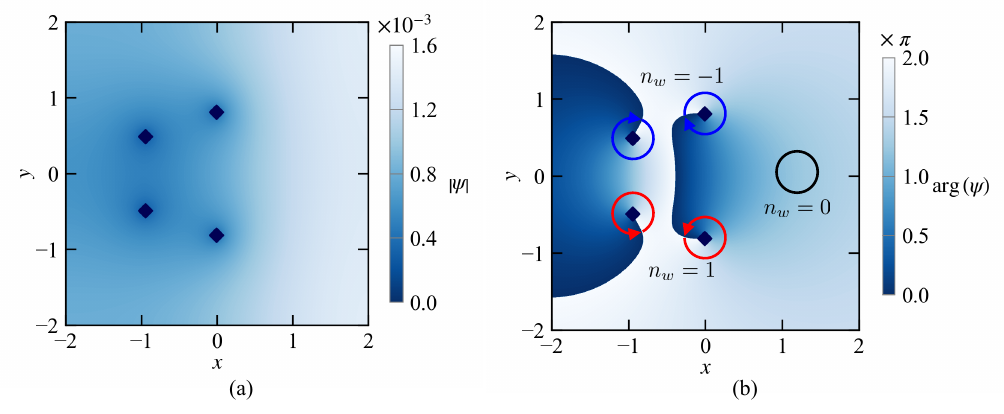}
    \caption{The distribution of (a) the magnitude and (b) the phase of a two-dimensional complex scalar field, representing a velocity field induced by four vortex elements as described by \eqref{eq:ghn}, with dark blue diamonds indicating the intersections of the real and imaginary parts of the complex scalar field.
   In (b), the red and blue circular arrows indicate regions where the winding number around the zero-level set of the complex scalar field is \(\pm 1\), while the black circle highlights areas with a winding number of 0.}
    \label{fig:winding}
\end{figure}

We consider a two-dimensional example to illustrate the representation of the flow field by the complex scalar field. Figure~\ref{fig:winding} shows the distribution of the magnitude and phase of the complex scalar field, which represents a velocity field induced by four vortex elements, as described by \eqref{eq:ghn}. In figure~\ref{fig:winding}(b), a closed curve \( \partial \Sigma \) is depicted in \( \mathbb{R}^2 \), with the red and blue circular arrows indicating regions where the winding number around the zero-level set of \( \psi \) is \(\pm 1\), and the black circle highlighting areas where the winding number is zero.

If the zero-level set of \( \psi \) does not intersect the surface \( \Sigma \) enclosed by \( \partial \Sigma \), the winding number, computed using \eqref{eq:winding}, is zero, implying the absence of net circulation or vorticity flux through the surface. Conversely, a non-zero vorticity flux, as given by \eqref{eq:ghn}, indicates that the zero-level set of \( \psi \) intersects the surface \( \Sigma \). In this case, the flow field is characterized by the presence of vortex filaments, which correspond to the zero-level set of \( \psi \) and represent coherent structures within the flow, where the complex scalar field \( \psi \) undergoes a significant phase change.

On the other hand, interpreting the winding number as the number of intersections between the vortex filaments and the surface \( \Sigma \) suggests that these filaments possess a uniform strength. This strength is represented by the constant \( h \), which links the circulation of the velocity field to the winding number. As a result, each vortex filament contributes equally to the total circulation, providing a simplified yet insightful characterization of the flow’s topological structure.

For a time-dependent, complex-valued level set function \(\psi\), assuming the function satisfies the necessary regularity conditions, its time evolution near the zero-level set is governed by the equation
\begin{equation}
\frac{\partial \psi}{\partial t} + \bm{u} \cdot \bm{\nabla} \psi = 0,
\label{eq:DpsiDt}
\end{equation}
where \(\bm{u}\) is a velocity field that governs the motion of the zero-level set of \(\psi\). This equation describes the advection of the implicit vortex filament, specifying how it is transported by the flow of the velocity field \(\bm{u}\). In particular, the evolution of the zero-level set, where \(\psi = 0\), follows the flow determined by \(\bm{u}\), thereby capturing the dynamics of the vortex filament. This leads to the following theorem.

\begin{theorem}[Relation between curve dynamics and implicit representation]
Let \(\psi\) be a time-dependent complex level set function that satisfies the regularity assumptions, namely, there exists a neighborhood \(\mathcal{U}\) of the zero-level set of \(\psi\) such that the gradient matrix of \(\psi\) is of full rank in \(\mathcal{U}\). Denote by \(\bm \gamma\) the set of vortex filaments advected by the velocity field \(\bm{u}_{\gamma}\), and let \(\psi\) represent the complex scalar field associated with these vortex filaments, as defined in definition \ref{defn:implicit}. The evolution of the zero-level set of \(\psi\) is governed by the \eqref{eq:DpsiDt}, where the velocity field \(\bm{u}\) and the relative velocity \(\bm{u}_{\gamma}\) are aligned in the direction normal to the filaments. This is expressed by the condition
\begin{equation}
(\bm{u} - \bm{u}_{\gamma}) \times \frac{\partial}{\partial s} \bm \gamma(s,:) = \bm{0}.
\end{equation}
\end{theorem}
The velocity components that primarily influence the evolution of the curves are those defined on the zero-level set. Additionally, the position of the curves remains unaffected when they are translated along their tangent direction, analogous to rotating a circle about its axis of symmetry. Thus, it is sufficient to constrain the normal and binormal components of the velocity field.

% 上面讲的对速度进行修正和约束，和我们后面要讲的东西有什么关系么？我总觉得有点突兀欸

% 要不要加一些连接之类的？

\section{Quantum methodology for implicit representation of vortex filaments}
\label{sec:overview}

\begin{figure}
    \centering
    \includegraphics[width=\textwidth]{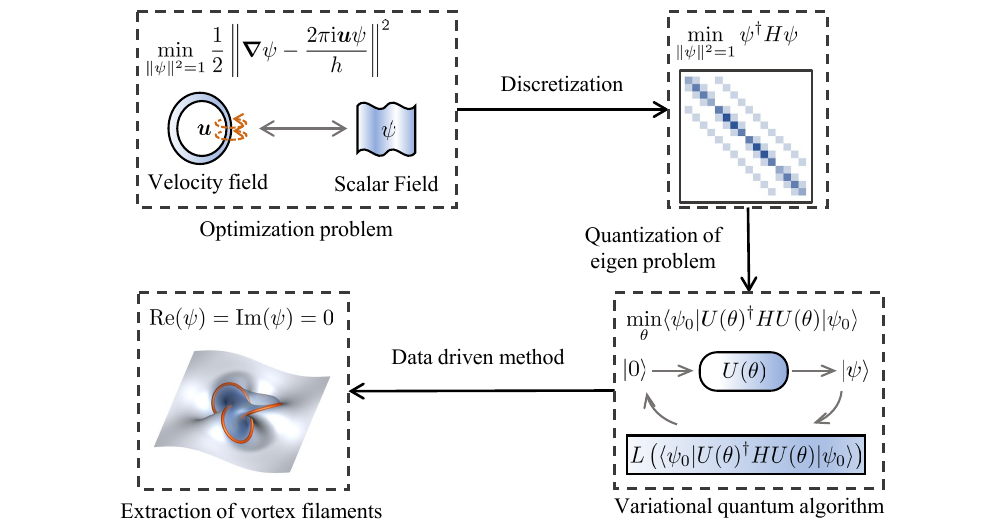}
         \caption{Schematic of the proposed methodology for preparing a quantum state and extracting vortex filaments from a velocity field. The process begins with the formulation of an optimization problem to establish a relationship between velocity circulation and the winding number of a complex scalar field. The problem is discretized on a staggered grid, leading to an eigenvalue problem for the minimum eigenvalue of a Hermitian matrix. A VQE is applied to compute the corresponding eigenvector, yielding a quantum state that encodes the velocity field data. Measurement of the quantum state and neural network processing then enable the extraction of vortex filaments, providing a low-dimensional representation of the velocity field.}
    \label{fig:procedures}
\end{figure}

As shown in figure~\ref{fig:procedures}, we propose a method for representing a given velocity field as a quantum state suitable for quantum computation, enabling the extraction of the corresponding vortex filaments. The method begins by formulating an optimization problem, based on the approximate linear relationship between the velocity circulation and the winding number of a complex scalar field, thereby facilitating the reconstruction of the scalar field from the velocity field as detailed in the previous section. This optimization is discretized on a staggered grid, leading to an eigenvalue problem for a Hermitian matrix. A VQE is subsequently applied to compute the eigenvector associated with the minimum eigenvalue, resulting in a quantum state that encodes the discrete velocity field. Finally, by measuring the quantum state and employing a neural network, vortex filaments are identified, providing a reduced-dimensional representation of the velocity field.

\subsection{Discretization of the optimization formulation}
\label{sectionDiscre}

We solve the optimization problem \eqref{eq:optpsi} on a staggered grid, where the complex scalar field is discretized at the cell centers and the velocity components are staggered at the face centers. The staggered grid arrangement, first introduced by \citet{Harlow1965numerical} in the marker-and-cell (MAC) method, places velocity components at cell faces while locating scalar quantities at cell centers. This geometric configuration provides dual numerical advantages: it inherently maintains discrete conservation laws for improved solution stability, while simultaneously enabling precise calculation of vorticity flux and velocity circulation that are crucial for vortex dynamics analysis.

\begin{figure}
    \centering
    \includegraphics[width=0.7\textwidth]{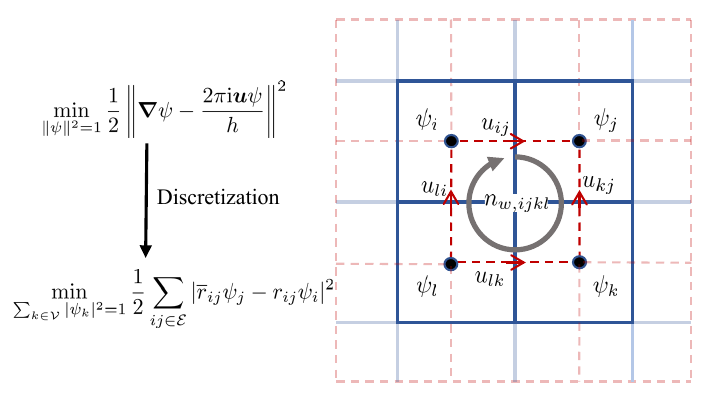}
    \caption{Discretization of the optimization problem \eqref{eq:optpsi} on a staggered grid, where the complex scalar field \( \psi \) is defined at the cell centers and the velocity \( \bm u \) is defined at the face centers. For adjacent cells \( i \) and \( j \), the velocity component \( u_{ij} \) at the shared face leads to the discretization given by \eqref{eq:psiij}. This formulation allows the optimization problem to be recast as the minimization of \eqref{eq:ijkpsi} over the sets of cells \( \mathcal{V} \) and faces \( \mathcal{E} \).}

    \label{fig:discretization}
\end{figure}

This discretization method follows the approach outlined by~\citet{Xiong2022Clebsch}. As shown in figure~\ref{fig:discretization}, for two adjacent cells, labeled \(i\) and \(j\), the complex values \( \psi_i \) and \( \psi_j \) are placed at the respective cell centers, while the shared face between the cells contains the velocity component \( u_{ij} \). On the \(ij\)-th face, the term \( \bm \nabla \psi - \mathrm{i} \hat{\bm u} \psi \) is discretized as
\begin{equation}
 (\psi_{j} - \psi_{i}) -  \frac{1}{2} \textrm{i} \hat \eta_{ij} ( \psi_{i} +  \psi_{j})+\mathcal{O}(\Delta x^2),
\label{eq:psiij}
\end{equation}
where \( \hat \eta_{ij} = 2\pi u_{ij} \Delta x / h \) and \( \Delta x \) is the grid cell spacing.

We apply a Taylor expansion to obtain
\begin{equation}
    1 \pm \frac{1}{2} \textrm{i} \hat \eta_{ij} = \exp\left(\pm \frac{\textrm{i}}{2} \hat \eta_{ij}\right) + \mathcal{O}(\Delta x^2).
\end{equation}
This enables the symmetric representation of \eqref{eq:psiij} as
\begin{equation}
\overline{r}_{ij}\psi_j - r_{ij}\psi_i + \mathcal{O}(\Delta x^2),
\label{eq:psiij2}
\end{equation}
where \( r_{ij} = \exp(\textrm{i} \hat{\eta}_{ij}/2) \), and \( \overline{r}_{ij} \) denotes the complex conjugate of \( r_{ij} \). The optimization problem can then be reformulated as
\begin{equation}
\min_{\psi_k,k\in \mathcal{V}} \frac{1}{2}\sum_{ij\in \mathcal{E}} |\overline{r}_{ij}\psi_j - r_{ij}\psi_i|^2~~\text{s.t.}~~ \sum_{k\in \mathcal{V}}|\psi_k|^2 = 1,
\label{eq:ijkpsi}
\end{equation}
where \( \mathcal{V} \) is the set of all cell indices and \( \mathcal{E} \) is the set of all face indices.

The term \( |\overline{r}_{ij}\psi_j - r_{ij}\psi_i|^2 \) can be expanded as
\begin{equation}
\left( \overline{ \overline{r}_{ij}\psi_j - r_{ij}\psi_i} \right) \left(\overline{r}_{ij}\psi_j - r_{ij}\psi_i\right)=|\psi_i|^2+|\psi_j|^2-(r_{ij})^2\overline{\psi}_j\psi_i-(\overline{r}_{ij})^2\overline{\psi}_i\psi_j
\end{equation}
Using this expansion, the optimization problem \eqref{eq:ijkpsi} can be reformulated as a quadratic form of complex variables under a normalization constraint
\begin{equation}
\min_{\|\psi\|^2=1}  \psi^\dagger H \psi
\label{eq:eigen}
\end{equation}
where the elements of the Hermitian matrix \( H \) are given by
\begin{equation}
    H_{ij} = 
    \begin{cases}
    - (\overline{r}_{ij})^2 & \text{if the } j\text{-th cell is a neighbor of the } i\text{-th cell}, \\
    d_i & \text{if } i = j, \\
    0  & \text{otherwise}.
    \end{cases}
\end{equation}
Here, \( d_i \) denotes the number of neighbors of the \( i \)-th cell and corresponds to the term \( |\psi_i|^2 \). The term \( -(\overline{r}_{ij})^2 \) represents the cross terms, specifically \( (r_{ij})^2 \overline{\psi}_j \psi_i \) and \( (\overline{r}_{ij})^2 \overline{\psi}_i \psi_j \). When \( i \) and \( j \) are not neighbors, \( H_{ij} = 0 \). Since \( r_{ij} = \overline{r}_{ji} \), the matrix \( H \) is Hermitian, transforming the filament extraction problem into a Hermitian eigenvalue problem.

We remark that in \eqref{eq:optpsi}, \( \psi \) represents a continuous complex scalar field defined over the fluid domain. In the discretized formulation \eqref{eq:eigen}, \( \psi \) is treated as a vector, with each component corresponding to the complex value of the scalar field at the center of a grid cell. This transition to a discrete representation is necessary for numerical computation.

\subsection{Quantum algorithm}
\label{sec:VQE_solve}

We represent the discrete scalar field \(\psi\) in \eqref{eq:eigen} as the quantum state \(|\psi\rangle\) and interpret the Hermitian matrix as the Hamiltonian operator. Thus, equation \eqref{eq:eigen} is reformulated as a quantum eigenvalue problem. To solve this, we employ a VQE.

\begin{figure}
    \centering
\begin{quantikz}
    \lstick{$\ket{0}$} & \gate{\mathrm{R}_y(\theta)} \gategroup[wires=5,steps=6,style={dashed, rounded corners}, label style={label position=below,anchor=north,yshift=-0.3cm}]{Repeated for $N_d$ times} & \gate{\mathrm{R}_z(\theta)} & \ctrl{1} & \gate{\mathrm{R}_y(\theta)} & \gate{\mathrm{R}_z(\theta)} &         & \, \hdots \,  & \meter{}  \gategroup[wires=5,steps=1,style={dashed, rounded corners}, label style={label position=below,anchor=north,yshift=-0.3cm}]{Measurements}\\
    \lstick{$\ket{0}$} & \gate{\mathrm{R}_y(\theta)} & \gate{\mathrm{R}_z(\theta)} & \targ{} & \gate{\mathrm{R}_y(\theta)} & \gate{\mathrm{R}_z(\theta)} & \ctrl{1} & \, \hdots \,  & \meter{} \\
    \lstick{$\ket{0}$} & \gate{\mathrm{R}_y(\theta)} & \gate{\mathrm{R}_z(\theta)} & \ctrl{1} & \gate{\mathrm{R}_y(\theta)} & \gate{\mathrm{R}_z(\theta)} & \targ{} & \,  \hdots \,  & \meter{}  \\
    \lstick{$\ket{0}$} & \gate{\mathrm{R}_y(\theta)} & \gate{\mathrm{R}_z(\theta)} & \targ{} & \gate{\mathrm{R}_y(\theta)} & \gate{\mathrm{R}_z(\theta)} & \ctrl{1} & \, \hdots \,  & \meter{} \\
    \lstick{$\ket{0}$} & \gate{\mathrm{R}_y(\theta)} & \gate{\mathrm{R}_z(\theta)} &         & \gate{\mathrm{R}_y(\theta)} & \gate{\mathrm{R}_z(\theta)} & \targ{} & \, \hdots \,  & \meter{}
\end{quantikz}
    \caption{Schematic of the parameterized quantum circuit \( U(\theta) \) for a system of \(N_\mathrm{qubit}\) qubits, with \(N_\mathrm{qubit} = 5\) used as an illustrative example. The circuit consists of \( N_d \) blocks, each containing four layers of rotational gates with distinct parameters. After every two layers of rotational gates, a layer of CNOT gates is applied to connect adjacent qubits.}
    \label{fig:circuit}
\end{figure}
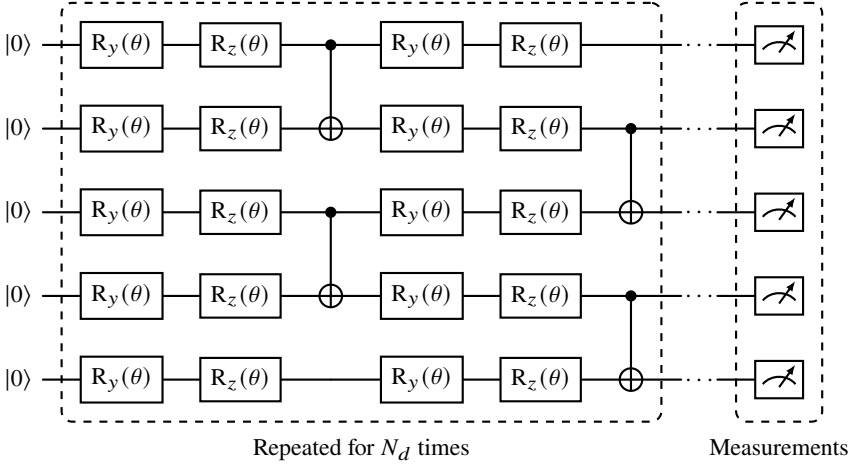

The quantum circuit \( U(\bm{\theta}) \) is parameterized by the vector \(\bm{\theta}\), representing the circuit parameters. The initial state is prepared as the zero state, \( |\psi_0\rangle = |0\rangle \). The structure of the circuit is shown in figure~\ref{fig:circuit}, where it consists of $N_\mathrm{qubit}$ qubits and $N_d$ basic blocks. The number of qubits $N_\mathrm{qubit}$ depends on the discretization of the computational domain. In our case, it's defined as
\begin{equation} 
    N_\mathrm{qubit} \sim \log_2 N_{\mathrm{cell}}.
\end{equation}
The blocks share identical structure but with different parameter values. Each block contains four layers of rotational gates, defined as
\begin{equation}
\begin{aligned}
\textrm{R}_y(\theta) = \left[\begin{matrix}\cos \left(\theta/2\right) & -\sin \left(\theta/2\right) \\ \sin \left(\theta/2\right) & \cos \left(\theta/2\right)  \end{matrix}\right],~~ \textrm{R}_z(\theta) = \left[\begin{matrix}\exp\left(-\mathrm{i} \theta/2\right) & 0 \\0 & \exp\left(\mathrm{i} \theta/2\right)   \end{matrix}\right],
\end{aligned}
\end{equation}
followed by a layer of CNOT gates connecting adjacent qubits after every two layers of rotational gates. While rotational gates play a crucial role in quantum circuits \citep{Xiao2024physics}, we restrict the use of rotational gates to $\textrm{R}_y$ and $\textrm{R}_z$ for two reasons. First, $\textrm{R}_x$ gate can be expressed by $\textrm{R}_y$ and $\textrm{R}_z$, which means including all three does not fundamentally expand the expressibility of the circuit. Second, our circuit structure is inspired by the hardware-efficient ansatz, where minimizing the variety of quantum gates enhances practical implementability on near-term quantum devices \citep{Kandala2017hardware, Huang2022robust}. To expand the search space, the block is further repeated \( N_d \) times. The effect of \( N_d \) on circuit performance is discussed in the following section.

The parameterized circuit \( U(\bm{\theta}) \) is applied to the initial state \( |\psi_0\rangle \), leading to the evolution of the system to the target state \( |\psi\rangle = U(\bm{\theta}) |\psi_0\rangle \). The expectation value of the Hamiltonian operator \( H \) is then computed as the objective function for optimization
$\langle \psi | H | \psi \rangle$,
where the parameters \( \theta \) are adjusted to minimize this function. The optimization procedure yields the solution 
\begin{equation}
\bm{\theta}_{\textrm{opt}} = \mathop{\arg \min} \limits_{\theta} \langle \psi_0 | U(\bm{\theta})^\dagger H U(\bm{\theta}) | \psi_0 \rangle,
\end{equation}
which results in the optimal state \( |\psi_{\textrm{opt}} \rangle = U(\bm{\theta}_{\textrm{opt}}) | \psi_0 \rangle \).

When measuring the objective function \(\langle \psi | H | \psi \rangle\) on a quantum computer, the Hamiltonian \( H \) is typically decomposed into a sum of Pauli operators as
\begin{equation}
H = \sum _{i} c_i P_i,
\label{eq:decompose}
\end{equation}
where \( P_i \) are tensor products of Pauli matrices and \( c_i \) are real coefficients. The objective function then becomes
\begin{equation}
\langle \psi | H | \psi \rangle = \sum_i c_i \langle \psi | P_i | \psi \rangle .
\label{eq:measurement}
\end{equation}
This decomposition allows for the Hamiltonian measurement to be simplified to the evaluation of individual Pauli terms \citep{Mcclean2016theory}. 

As the system size increases, the number of Pauli operators grows, leading to a significant increase in measurement costs. To address this, the ``Pauli operator cutoff" method is adopted, which reduces the number of measurements while maintaining an appropriate balance between efficiency and accuracy. Further details of this approach are provided in appendix~\ref{sec:PauliCutoff}.

\subsection{Extraction of vortex filaments}
\label{sec:extract}

The computed quantum state obtained using VQE serves as the basis for identifying the locations of vortex filaments. To determine whether a filament intersects a given surface, the winding number around a closed curve or surface element is calculated by evaluating \eqref{eq:winding}. A non-zero winding number indicates the presence of a vortex filament, whereas a zero value suggests no intersection. If a filament is detected, interpolation is employed to refine the location, accurately pinpointing the crossing point within the discrete grid.

On a discrete grid, we assume that the centers of four adjacent cells, denoted \( jklm \), form a square with side length \( \Delta x \). The winding number over this square is approximated by \citet{Weibmann2014}
\begin{equation}
    n_{w,jklm} = \frac{1}{2\pi}\left[\arg \left(\frac{\psi_k}{\psi_j} \right) + \arg \left(\frac{\psi_l}{\psi_k} \right) + \arg \left(\frac{\psi_m}{\psi_l} \right) + \arg \left(\frac{\psi_j}{\psi_m} \right) \right],
    \label{eq:DiscreWinding}
\end{equation}
where \(\psi_i\) (with \(i = j, k, l, m\)) denotes the discretized complex scalar field at the grid points corresponding to \(j\), \(k\), \(l\), and \(m\). If \( n_{w,jklm} = 0 \), no vortex filament is considered to intersect the square formed by the four adjacent cells \( jklm \). If \( n_{w,jklm} \neq 0 \), it indicates the presence of a vortex filament intersecting the square. In such cases, the exact crossing point (or zero point) is determined using linear interpolation, given by
\begin{equation}
    (1 - v)(1 - u) \psi_j + u(1 - v) \psi_k + (1 - u) v \psi_m + u v \psi_l = 0,
    \label{eq:interpolation}
\end{equation}
where \( (u, v) \) are local coordinates within the unit square \((0, 1) \times (0, 1)\), representing the relative position of the intersection of the vortex filament within the square. Solving this equation yields the precise location of the intersection.

Although theoretically effective, quantum computations face practical challenges due to noise and errors. Quantum decoherence, caused by interactions with the environment, can drive a quantum state toward relaxation to the ground state $|0\rangle$, and quantum gates may introduce gate errors during operations \citep{Bharadwaj2024Simulating}. Other sources, such as limited grid resolution, measurement inaccuracies, and incomplete convergence from a finite number of iterations in the quantum algorithm, also introduce noise into the computed quantum state. This noise complicates the reliable identification of vortex filaments.

% 新加的内容：我们使用的神经网络有两个分支：一个分支预测网格面上是否有涡丝穿过，类似之前提到的winding number；另一个分支预测涡丝穿过位置的局部坐标。两个分支作为一个整体进行训练，并在loss function中同时包含了flux conservation、symmetry等因素，在提高准确性的同时保证vortex filament的物理性质。
To mitigate these challenges, we propose a data-driven machine learning approach for extracting vortex filaments from noisy quantum states. Our neural network comprises two branches: one branch predicts whether a vortex filament crosses the grid surface, analogous to the discretized winding number discussed in equation~\eqref{eq:DiscreWinding}, while the other branch forecasts the local coordinates of the crossing location. The two branches are trained collaboratively, with the loss function incorporating factors such as flux conservation and symmetry, thereby enhancing precision and preserving the physical properties of the vortex filaments. The circuit depth in training and testing dataset is intentionally reduced to simulate scenarios with noise and error. This method augments the classical interpolation technique, improving resilience to noise and enhancing detection accuracy. By learning patterns in noisy data, the machine learning model can identify filament locations even when classical methods are compromised by noise artifacts. This data-driven approach leverages the robustness of classical algorithms while exploiting the adaptability of machine learning, ensuring accurate and efficient vortex filament extraction in complex, noisy environments. Further details of the machine learning method and its implementation are provided in appendix~\ref{sec:NNextraction}.

\begin{figure}
    \centering
    \includegraphics[width=10cm]{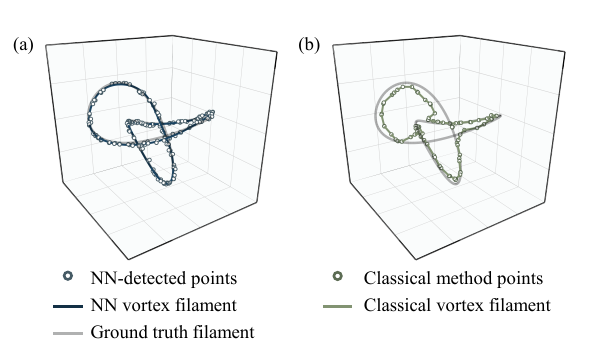}
    \caption{(a) The vortex filament using the proposed neural network method exhibit near-perfect agreement with the target filament locations. NN denotes neural network. (b) In contrast, the extracted vortex filaments using classical method show significant deviation from the ground truth.}
    \label{fig:NN and Cla}
\end{figure}

The effectiveness of the proposed method is demonstrated in figure~\ref{fig:NN and Cla}, where vortex filaments reconstructed using the network approach exhibit near-perfect alignment with the true filament locations. In contrast, filaments extracted using the classical method show significant deviations from the ground truth. This comparison underscores the superior accuracy of the machine learning-based method, particularly in the presence of noise.

% The issue arises because the 

\section{Assessment of quantum implicit representation}

%量子表征精度的评估-熊
%\section{Algorithm configuration}
\label{sec:implement}
We provide a comprehensive analysis of the method's computational complexity and configuration of key parameters, including the quantum circuit depth \(N_d\), the number of iterations \(N_i\), the vortex strength \(h\) and the core size \(a\), to evaluate both the potential acceleration and the robustness of the proposed method.

\subsection{Parameter configuration}
\label{sec:sensitivity}
We analyze the influence of different parameters and identify the most suitable configurations. One challenge lies in the direct comparison of the extracted vortex filaments with the target ones, especially in complex flow fields where defining positional discrepancies is nontrivial. Alternatively, considering our goal of establishing a connection between the vortex filaments and the velocity field, we reconstruct the velocity field using the RM model referred in equation~\eqref{eq:RM}. We then quantify the discrepancy between the recovered velocity field and the original velocity field using the mean squared error (MSE) metric:
\begin{equation}
    \mathrm{MSE} = \frac{1}{N_{\text{cell}}} \sum_{\text{cells}} \left|\boldsymbol{u} - \boldsymbol{u}_{\text{target}}\right|^2,
    \label{eq:mse}
\end{equation}
where \(N_{\text{cell}}\) denotes the total number of grid cells in the flow field. This metric provides a quantitative assessment of the accuracy of the reconstructed flow field, offering insights into the algorithm’s effectiveness in capturing the dynamics of the system.

We inventory the parameters that require analysis. Figure~\ref{fig:circuit} illustrates the structure of the quantum circuit, where a single block is repeated \(N_d\) times. Additionally, the VQE requires iterative optimization, and the number of iterations, \(N_i\), is another parameter of interest. Since both parameters are closely related to the optimization process, we refer to them collectively as optimization parameters. Meanwhile, the vortex strength parameter \(h\), introduced in equation~\eqref{eq:ghn}, and the vortex core size \(a\) in the RM model \eqref{eq:RM} are grouped as vortex parameters, as they illustrate the vortex characteristics.

The choice of optimizer and hyperparameter settings also significantly impact the performance. In this study, we use the Adam optimizer with an initial learning rate of 0.01, which decays by a factor of 0.99 every 10 steps. These parameters ensure stability and fast convergence during optimization. While a systematic exploration of these hyperparameters' effects is beyond the scope of this study, preliminary experiments suggest that they are reasonable choices under the proposed method. 

We perform a detailed analysis of the two parameter categories. The optimization parameters include the circuit depth $N_d$ and the iteration number $N_i$. The circuit depth $N_d$ determines the scale of the quantum circuit, and in the context of VQE, it defines the search space for $\psi$. A deeper circuit can lead to better results but may become less effective due to increased noise and slower convergence \citep{Cerezo2021variational}, while also demanding more parameters, further hindering optimization efficiency. Similarly, the iteration number $N_i$ controls the optimization steps: more iterations tend to improve results and robustness but at the cost of increased computational time \citep{Lavrijsen2020Classical}.

\begin{figure}
    \centering
    \includegraphics[width=\linewidth]{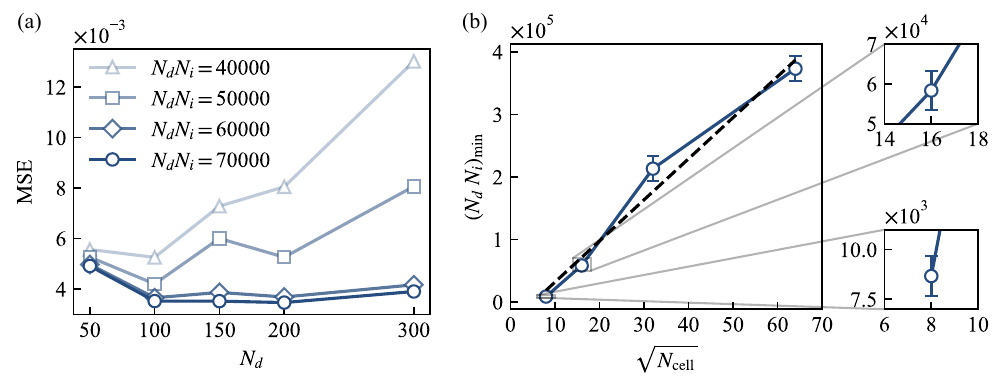}
    \caption{Results of the optimization parameter study. (a) Performance on a \(16 \times 16\) grid for varying complexities and circuit depths, demonstrating that complexity has a dominant influence, with optimal depths in the range of 100--200. (b) Estimated minimum complexity $\left(N_d N_i \right)_{\min}$ required for convergence as a function of $\sqrt{N_\mathrm{cell}}$, showing a linear trend.}
    \label{fig:complexity}
\end{figure}

The experiments were conducted using the ``frog leaping'' example, shown in figure~\ref{fig:winding}. Initially, we discretized the case into a $16 \times 16$ grid to investigate the relationship between $N_i$, $N_d$, and the mean squared error (MSE).  As shown in figure~\ref{fig:complexity}(a), the product of $N_i$ and $N_d$ dominates performance. We refer to this product as the complexity of the VQE, as the number of quantum gates is proportional to $N_d$, and each gate must be updated $N_i$ times. Along the horizontal axis, increasing $N_d$ reduces MSE significantly at first, reaching optimal performance within a specific range ($N_d = 100$ to $N_d = 200$), after which performance stabilizes or deteriorates. Along the vertical axis, higher complexity ($N_i N_d = 60000$ or $70000$) achieves lower and more stable MSE.
The interaction between $N_d$ and $N_i N_d$ indicates that while complexity dominates performance, selecting an appropriate $N_d$ within the optimal range is crucial for efficient optimization, highlighting the trade-off between resource allocation and performance.

To further investigate the role of VQE complexity, we estimate the minimum required for convergence across different grid resolutions, ensuring that \(N_d\) and \(N_i\) remain within an appropriate range for effective optimization. Figure~\ref{fig:complexity}(b) illustrates a linear relationship between the minimum complexity \((N_d N_i)_{\min}\) and the grid dimension, given by \(\sqrt{N_\text{cell}}\). The problem is discretized into grids of \(8 \times 8\), \(16 \times 16\), \(32 \times 32\), and \(64 \times 64\), with optimal \(N_d\) values of 150, 200, and 250 for all the first three smaller grids, and 2000, 2200, and 2400 for the \(64 \times 64\) grid due to increased scale of the problem. The results are averaged over three different values of \( N_d \) for each grid to obtain representative \((N_d N_i)_{\min}\) values, as shown in the zoomed-in views. The observed linear trend in the plot suggests that the minimum complexity required for convergence is proportional to \(\sqrt{N_\text{cell}}\), offering a practical guideline for estimating computational costs as problem size increases. We found that the product $N_d N_i$ is the primary determinant for VQE performance. Moreover, we identified a linear relationship between the minimum complexity and the grid resolution, providing valuable insights for optimizing VQE parameters.

The vortex parameters considered in this study include the vortex strength $h$ introduced in \eqref{eq:ghn} and the vortex core size $a$ introduced in \eqref{eq:RM}. In our mathematical model, $h$ represents an estimate of the actual vortex strength. This parameter is straightforward to determine in simplified cases, such as the ``frog leaping'' example described earlier, where all vortices share identical strength, and the velocity field is computed using the BS or RM model. However, in more complex cases like turbulence, accurately estimating vortex strengths becomes significantly more challenging, as the vortices exhibit varying intensities. For instance, velocity and energy spectra are commonly employed to characterize turbulence \citep{Kolmogorov1941Local, Lundgren1982Strained}. Moreover, a strong vortex can be approximated as the superposition of weaker vortices when $h$ is reduced, but our grid-based approach constrains each grid cell to a single vortex filament. The trade-off between these factors merits further exploration.

The core size $a$ not only prevents singularities but also captures variations in vortex structure, which is essential for accurately modeling more complex vortex behaviors. For instance, in the ``frog leaping'' experiments by \citet{Lim1997Note}, \citet{Jerrard2018Leapfrogging}, and \citet{Wang2024Deformation}, even idealized point vortices exhibit deviations from expected behavior. Consequently, selecting appropriate values for $h$ and $a$ is critical for accurate velocity reconstruction.

%为了验证两个参数的真正表现，我们使用了一个湍流算例来进行研究。湍流算例的具体描述请看后面。然后我们发现对于h来说，越细的网格所需要的h越小，证明我们之前对于叠加的论述是对的。同时，我们发现越细的网格所需要的eps也越小。通过上面的一系列实验我们发现，在更加细的网格当中，表现最好的$h$和$a$一般都会更好。一方面，这个结果印证了我们之前对于强涡可以使用弱涡叠加的论断：更细的网格会有更多的网格数量，也就允许更多的涡存在，更好支持这种叠加。另一方面，对于$a$的下降，我们也猜测在更加细的网格当中，每个涡的核大小都会更小，更多的涡也会自带一些smoothing的效果。

\begin{figure}
    \centering
    \includegraphics[width=\linewidth]{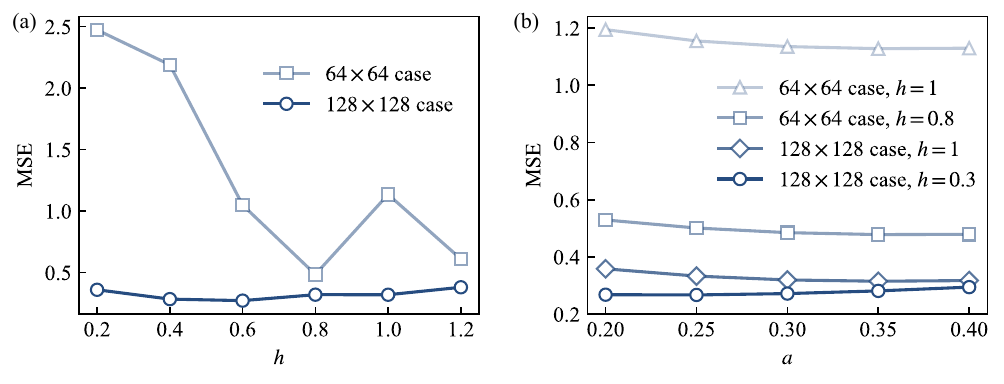}
    \caption{Results of the vortex parameter investigation. (a) The impact of \( h \) across different grid resolutions. In the \( 64 \times 64 \) grid, the optimal value of \( h \) is 0.8, while in the \( 128 \times 128 \) grid, it decreases to 0.3. \(a\) is fixed at 0.3 (b) The effect of the core size \( a \). The optimal \(a\) for \( 64 \times 64 \) case \(h = 0.8\) is 0.35, and for \( 128 \times 128 \) case \(h = 0.3\) is 0.25, showing a consistent reduction in the optimal core size \(a\) as grid cell size decreases across the cases.}
    \label{fig:vortex_param}
\end{figure}

To evaluate the influence of these parameters in complex cases, we conducted experiments using a turbulent velocity field, with detailed descriptions provided in Section~\ref{sec:Turbulence}. The case was discretized on $64 \times 64$ and $128 \times 128$ grids to assess performance under varying resolutions. As shown in figure~\ref{fig:vortex_param}(a),  with $a$ fixed at 0.3, the optimal \( h \) value is relatively larger in coarser grids, while the optimal value decreases in finer grids. This indicates that higher grid resolutions tend to favor smaller vortex strengths, aligning with our hypothesis regarding vortex superposition. Figure~\ref{fig:vortex_param}(b) shows the relationship between the core size \( a \) and the mean squared error (MSE) for different grid resolutions and vortex strengths. MSE generally decreases as \( a \) increases, reaching a minimum before slightly increasing at larger \( a \). Based on these experiments, we observe that the optimal values of \(h\) and \(a\) in finer grids are generally smaller in value. On the one hand, this result supports our previous assertion that strong vortices can be effectively represented by the superposition of weaker vortices: finer grids contain more cells, allowing for more vortices and better supporting the superposition. On the other hand, regarding the decrease in \(a\), we speculate that in finer grids, the core size of each vortex becomes smaller, and the presence of more vortices inherently introduces a smoothing effect.

\subsection{Complexity analysis}
We analyze the computational complexity of our proposed approach. As outlined in section \ref{sec:overview}, the pipeline comprises three main stages: formulation of the optimization problem, the VQE, and post-processing. The latter two stages are computationally intensive and operate in consecutive order. We conduct a detailed complexity analysis of these two components.

% VQE使用一个含参量子电路生成波函数psi，测量目标函数的值之后，使用经典优化器对每个含参量子门的参数进行优化，经过N_i次迭代后得到更好的参数与psi。
The VQE employs a parameterized quantum circuit to prepare a trial scalar field $\psi$. The objective function $\langle \psi | H | \psi \rangle$ is then measured, according to which the parameters of each controlled quantum gate are optimized using classical optimizers. After $N_i$ iterations, an optimized set of parameters and the corresponding $\psi$ are obtai  ned.

% 计算复杂度的主要来源是优化中每次迭代中都要进行的反向传播（cite，参考suhao）。在每次迭代中，每个门都要根据梯度进行一次更新。在我们的问题中，对于某个大小的网格，有……大小的矩阵，对应……的qubits。根据我们之前的设计，一个basic block有4 * qubit个controlled gates，因此对于深度为N_d的量子电路，我们一次迭代需要更新……次。对于迭代N_i次的情况，则总共需要更新……次。
The primary computational cost of the VQE arises from back-propagation during each optimization step \citep{Cerezo2021variational}. In each iteration, every gate undergoes a gradient-based update. For a grid with $N_\mathrm{cell}$ cells, the problem involves a Hermitian matrix of size $N_\mathrm{cell} \times N_\mathrm{cell}$, corresponding to $\log(N_\mathrm{cell})$ qubits. According to our circuit design, a basic block in figure~\ref{fig:circuit} consists of $4 \log(N_\mathrm{cell})$ controlled gates (e.g., $R_y$ and $R_z$). In a quantum circuit with depth $N_d$, the total number of controlled gates becomes $4 N_d \log(N_\mathrm{cell})$. The execution time of the quantum circuit is porpotional to the circuit depth $N_d$. Thus, the time complexity of one optimization step can be expressed as

\begin{equation}
    T(N_\mathrm{cell}, N_d, N_i)_\mathrm{one\ step} \sim \mathcal{O}(N_d^2 \log N_\mathrm{cell}).
\end{equation}

For $N_i$ iterations, the overall complexity of the VQE is

\begin{equation}
    T(N_\mathrm{cell}, N_d, N_i)_\mathrm{VQE} \sim \mathcal{O}(N_d^2 N_i \log N_\mathrm{cell}).
\end{equation}

% 注意到N_i N_d 就是我们上一个part里面提到的VQE complexity. 我们在上面总结了minimum complexity大概和某某有线性关系，因此我们推测VQE优化总体的复杂度大概为……
The term $N_d N_i$ corresponds to the complexity discussed in \ref{sec:sensitivity}. Our analysis indicates that the minimum complexity $N_d N_i$ required for convergence scales linearly with the square root of the total grid size, which is approximately $\sqrt{N_\mathrm{cell}}$. Accordingly, we conjecture that the overall complexity of the VQE can be approximated as 
\begin{equation}
    T(N_d, N_\mathrm{cell})_\mathrm{VQE} \sim \mathcal{O}(N_d \sqrt{N_\mathrm{cell}} \log N_\mathrm{cell}).
\end{equation}
Since $\mathcal{O}(\sqrt{N_\mathrm{cell}} \log N_\mathrm{cell})$ is smaller than $\mathcal{O}(N_\mathrm{cell})$, the VQE complexity can be further bounded as
\begin{equation}
    T(N_d, N_\mathrm{cell})_\mathrm{VQE} \sim \mathcal{O}(N_d N_\mathrm{cell}).
\end{equation}

% 还有需要注意的一个点就是我们的post-process都需要测量完整的psi，可能对测量有要求。这个我之前没有提，现在也没有提，不知道应不应该保留。

% 对于post-process，我们在2.4（旧）中提到的经典后处理主要包含winding number计算和插值得到具体位置两步。
The classical post-processing method described in \ref{sec:extract} involves two main steps: computation of the winding number and interpolation.
% 对于winding number，每个面都需要进行一次计算，所以对于一个ijk的网格，如果假设ijk都是1/3次的话，那么共有这么多面，因此计算winding的复杂度为……。
For the winding number computation, in an $m \times j \times k$ grid, the discretized winding number must be evaluated for a total of $(m - 1)(j - 1)k + (m - 1)j(k - 1) + m(j - 1)(k - 1)$ faces. Assuming the grid dimensions \( m, j, k \) scale as \( {N_\mathrm{cell}}^{1/3} \), if the winding number must be computed for every face in the grid and each face requires a constant amount of computation, the time complexity of the winding number calculation can be approximated as:
\begin{equation}
    T(N_\mathrm{cell})_\mathrm{winding} \sim \mathcal{O}(N_\mathrm{cell}).
\end{equation}
% 插值只需要对winding不为0的进行。在最坏的情况下，我们假设每个面都需要插值，则插值的复杂度为：
The interpolation step is required only for faces where the winding number is non-zero. In the worst-case scenario, interpolation is performed on every face as described in \eqref{eq:interpolation}, yielding a total complexity of
\begin{equation}
    T(N_\mathrm{cell})_\mathrm{interpolation} \sim \mathcal{O}(N_\mathrm{cell}).
\end{equation}
% 后处理的总体复杂度为
Consequently, the overall complexity of the classical post-processing method is given by
\begin{equation}
    T(N_\mathrm{cell})_\mathrm{post} \sim \mathcal{O}(N_\mathrm{cell}) + \mathcal{O}(N_\mathrm{cell}) \sim \mathcal{O}(N_\mathrm{cell}).
\end{equation}
In addition, the data-driven post-processing approach exhibits a comparable complexity of $\mathcal{O}(N_\mathrm{cell})$, with further discussion provided in appendix \ref{sec:NNextraction}.

In summary, the overall complexity of the proposed pipeline is

\begin{equation}
    T(N_d, N_\mathrm{cell}) \sim \mathcal{O}(N_d \sqrt{ N_\mathrm{cell}} \log (N_\mathrm{cell})) + \mathcal{O}(N_\mathrm{cell}) \sim \mathcal{O}(N_d N_\mathrm{cell}).
\end{equation}

This complexity is considered advantageous. 
%According to our experiment setup, $N_d$ is much smaller than $N_\mathrm{cell}$, which means our complexity $O(N_d N_\mathrm{cell})$ outperforms the quadratic complexity $O(N_\mathrm{cell})^2$. 
We compare our method with several classical solvers commonly used for linear systems, including QR 
(Q for the orthogonal or unitary matrix and R for the upper triangular matrix ) decomposition, Gaussian elimination, and the Lanczos algorithm. QR decomposition and Gaussian elimination are direct methods, typically requiring $\mathcal{O}(N_\mathrm{cell}^3)$ for an $N_\mathrm{cell} \times N_\mathrm{cell}$ matrix, which makes them computationally expensive for large systems or fine grids \citep{Higham2011gaussian, Francis1961qr}. Iterative methods, such as Lanczos methods, can reduce complexity to $\mathcal{O}(m N_\mathrm{cell}^2)$ where $m$ denotes the number of desired eigenvalues \citep{Lanczos1950iteration}. 
and although only the minimum eigenvalue is required in our case, the computational cost remains quadratic at \( \mathcal{O}(N_{\mathrm{cell}}^2) \).
In contrast, our quantum approach achieves a complexity of $\mathcal{O}(N_d N_\mathrm{cell})$ under ideal conditions, with $N_d$ much smaller than $N_\mathrm{cell}$, offering a potential speedup.

However, it is important to note that the measurement of both the objective function \( \langle \psi | H | \psi \rangle \) and the final quantum state \( | \psi \rangle \) can be time-consuming, which was not initially considered in our analysis as the simulations were conducted on a virtual quantum machine with full access to state information. As described in Section~\ref{sec:VQE_solve}, the objective function is measured via decomposition of the Hamiltonian into Pauli operators. For an \( N_\mathrm{cell} \times N_\mathrm{cell} \) Hermitian matrix, this results in \( \mathcal{O}(N_\mathrm{cell}^2) \) distinct Pauli terms, leading to a measurement complexity of \( \mathcal{O}(N_\mathrm{cell}^2) \). To mitigate this overhead, we introduce a Pauli operator cutoff strategy in Appendix~\ref{sec:PauliCutoff} that does not require additional qubits. This strategy selectively removes Pauli terms with negligible contributions, thereby offering a simple and effective way to reduce the measurement burden while preserving noise-robust implementation. Another promising alternative is block-encoding, which embeds a non-unitary Hermitian matrix into a higher-dimensional unitary operator \citep{gilyen2019quantum, martyn2021grand}. This approach circumvents explicit Pauli decomposition and can be used in conjunction with the Hadamard test for efficient expectation value estimation. Nevertheless, it requires additional ancilla qubits and more complex quantum circuits, increasing susceptibility to noise, particularly in large-scale fluid simulations. In future work, block-encoding could be explored as a complementary technique within our framework.

In addition, the extraction of filaments from $\psi$ requires full knowledge of the final quantum state \( | \psi \rangle \). On actual quantum hardware, this necessitates quantum state tomography~\citep{o2016efficient}, which reconstructs the quantum state from multiple measurement outcomes. The corresponding measurement complexity scales as \( \mathcal{O}(N_\mathrm{cell} / \varepsilon^2 ) \), where \( \varepsilon \) denotes the target precision. To alleviate this challenge, we discuss two potential strategies. First, rather than reconstructing the entire quantum state, one may extract only essential information using techniques such as classical shadow tomography~\citep{huang2020predicting} and sparse tomography~\citep{chen2024enabling}, reducing the measurement complexity to \( \mathcal{O}(\log N_\mathrm{cell} / \varepsilon^2) \) under suitable assumptions. 
%Additionally, the numbering of computational cells can be optimized to facilitate these methods. 
Second, as introduced in Appendix~\ref{sec:NNextraction}, the machine learning-based post-processing framework can incorporate quantum noise and measurement uncertainty into the training process. This data-driven approach may relax the required measurement precision \( \varepsilon \), thereby further improving overall efficiency.

% In real quantum hardware, measurements follow quantum mechanical principles, providing probabilistic state information while being influenced by environmental interactions that introduce noise and decoherence, affecting outcomes and necessitating error-mitigation techniques. The computational cost of extracting useful information depends on the chosen measurement strategy, and the development of more efficient techniques remains an active area of research.

%复杂流场的量子隐式表征与低维结构提取
\section{Application of quantum implicit representation in turbulent flows}
To evaluate the capabilities of our method, we perform tests on four representative cases: two-dimensional point vortices, two-dimensional turbulence, three-dimensional knotted vortices, and three-dimensional turbulent flows. All quantum circuit simulations are performed using MindQuantum version 0.10.0 \citep{xu2024mindspore}. During the circuit parameter optimization phase, we compute the exact expectation values of observables using the statevector simulator, without sampling-based measurement. This approach is commonly adopted in numerical studies of VQAs to evaluate their idealized performance \citep{grimsley2019adaptive}.

\label{sec:results}
\subsection{Two-dimensional point vortices}
\begin{figure}
    \centering
    \includegraphics[width=\linewidth]{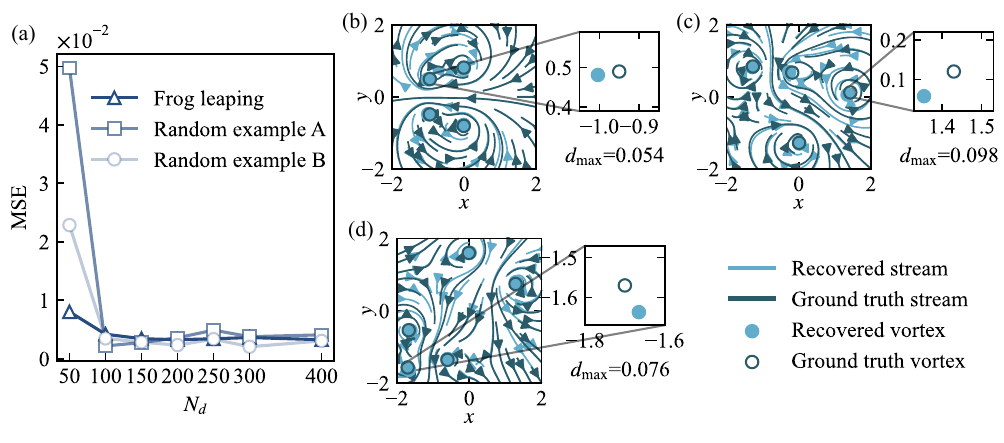}
    \caption{Evaluation of quantum circuit expressibility with three vortex configurations. (a) MSE converges below 0.01 for all cases. (b) The ``Frog Leaping'' case features a symmetric arrangement of two positive and two negative vortices. (c) ``Random Example A'' and (d) ``Random Example B'' are randomly generated systems with four and five vortices, respectively.}
    \label{fig:expressibility}
\end{figure}

We design three different two-dimensional vortex cases to test our method's ability to recover vortex dynamics. For all configurations, the circuit depth (\(N_d\)) ranges from 50 to 400, and each vortex configuration is discretized on a \(16 \times 16\) grid to accurately represent the flow field. Optimization is performed using the Adam optimizer with \(N_i = 300\) iterations to ensure model convergence. As shown in figure~\ref{fig:expressibility}(a), the MSE for all configurations converges below 0.01, demonstrating that the quantum circuits accurately capture the vortex dynamics. Specifically, the first, labeled ``Frog Leaping'' as shown in figure~\ref{fig:expressibility}(b), features a symmetric arrangement of two positive and two negative vortices, chosen to test the circuit's ability to capture well-defined, symmetric vortex patterns. The recovered streamlines closely match the ground truth, with a maximum distance error \(d_{\text{max}}\) of 0.054, indicating that the method performs well in handling simple and symmetric vortex systems.  The second and third configurations, ``Random Example A'' and ``Random Example B,'' are randomly generated systems with four and five vortices, respectively, as shown in figures \ref{fig:expressibility}(c) and \ref{fig:expressibility}(d). These setups challenge the method’s ability to capture more complex vortex distributions with varying strengths and spatial arrangements. In both cases, the recovered streamlines closely follow the general flow structure with slight deviations, and the recovered vortices align well with the ground truth, demonstrating the method's accuracy in tracking vortex positions despite minor discrepancies. 

% 实验结果表明，方法能够有效捕捉点涡的核心结构，验证了其在简单静态场景下的可靠性。

% 进一步地，为检验方法在含时过程中的表现，我们对典型的蛙跳运动（frog leaping motion）进行了实验。结果显示，该方法能够准确捕捉点涡在演化过程中的关键特征，展现了其在动态场景下的潜力。

\begin{figure}
    \centering
    \includegraphics[width=\linewidth]{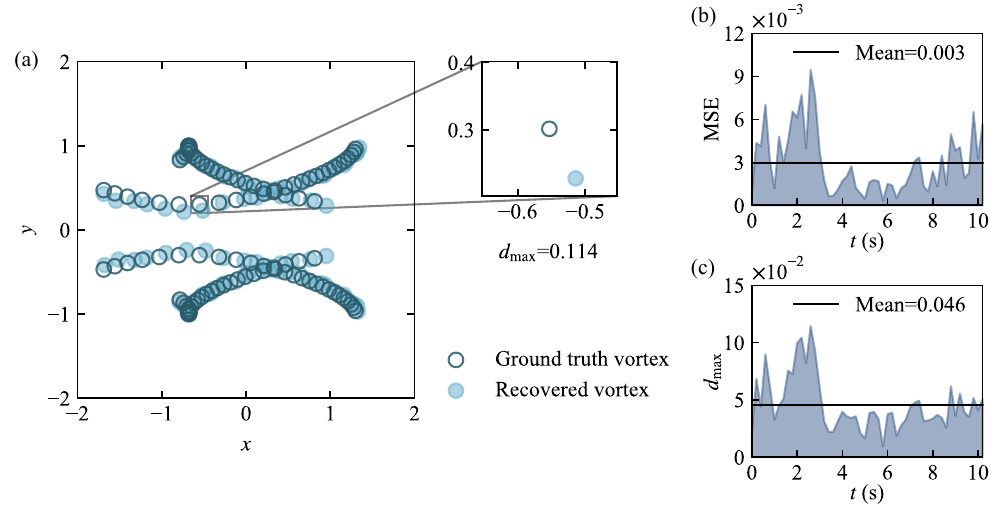}
    \caption{Evolution of ``Frog Leaping'' vortices. (a) Target and reconstructed vortices at selected time steps. The inset marks the maximum error region. (b, c) Time histories of maximum error ($d_{\max}$) and MSE, showcasing the method's performance throughout the simulation.}
    \label{fig:evolution}
\end{figure}

We also assess our method’s ability to capture the time-dependent ``Frog Leaping'' evolution, where four vortices in two pairs exhibit a periodic leapfrog-like motion. Driven by mutual advection, each vortex alternately leaps over and pursues the others, resulting in structured yet dynamic motion with periodic accelerations and decelerations. In numerical implementation, we divide the evolution into 50 time steps over a 10-second period, capturing the vortex behavior at regular intervals. The calculations are performed on a $32 \times 32$ grid within a quantum circuit of depth 250, optimized over 1200 iterations to ensure computational precision. The results, shown in figure~\ref{fig:evolution}, demonstrate the method's capability to accurately track the evolving vortex motion. With an average MSE of 0.003 and a maximum error of 0.046, the results reflect the method's robustness in characterizing the vortex dynamics. By accurately modeling the frog leaping motion, our method proves to be a valuable tool for studying and understanding vortex evolution in various fluid dynamics scenarios.

\subsection{Two-dimensional turbulent flows}
\label{sec:Turbulence}
To evaluate the applicability of the method to more complex scenarios, it is applied to two-dimensional turbulent flows. The velocity field is generated using the energy spectrum \(E(k) = \exp(-\alpha k^2)\), where \(k = \sqrt{k_x^2 + k_y^2}\) is the wavenumber, representing the magnitude of the wavevector in Fourier space, and \(\alpha\) is the decay factor that governs the energy distribution across different wavenumbers. A higher $\alpha$ leads to a faster decay of energy, favoring smaller-scale structures. The Fourier coefficients for the velocity components are given by:
\begin{equation}
\hat{u}(k) = -\frac{k_y}{k} E(k) e^{i \theta(k)}, \quad \hat{v}(k) = \frac{k_x}{k} E(k) e^{i \theta(k)},
\end{equation}
where $\hat{u}(k)$ and $\hat{v}(k)$ are the Fourier components of the velocity field in the \(x\)- and \(y\)-directions, respectively,
with \(\theta(k)\) representing random phases for isotropy. 
The inverse Fourier transform reconstructs the physical velocity field, converting the spectral representation into real-space velocity components and generating a spatially coherent turbulent structure. The flow intensity is controlled by scaling the velocity components with a factor \(A_{\mathrm{scale}}\). In the present computations, \(\alpha = 0.2\), \(A_{\mathrm{scale}} = 8000\), and the random seed is set to 7 for reproducibility. The flow is discretized on a \(128 \times 128\) grid with grid spacing \(\textrm{d}x = 0.6\). Optimization is performed with a circuit depth of \(N_d = 3000\) and \(N_i = 1333\) iterations.

\begin{figure}
    \centering
    \includegraphics[width=\linewidth]{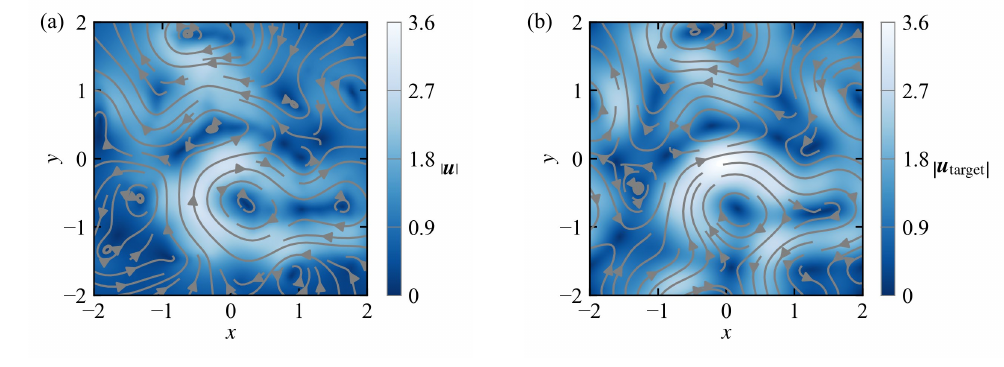}
    \caption{Results for two-dimensional turbulent flows. (a) The recovered velocity field \(\bm{u}\), with color representing velocity magnitude and gray streamlines indicating the flow direction, and (b) those of the target velocity field \(\bm{u}_{\mathrm{target}}\). }
    \label{fig:turbulence}
\end{figure}

The simulation results presented in figure~\ref{fig:turbulence} illustrate the complex turbulent structure and demonstrate the effectiveness of the method in dynamic flow scenarios. Figure~\ref{fig:turbulence}(a) shows the recovered velocity field \(\bm{u}\), with color representing velocity magnitude and gray streamlines indicating the flow direction, while figure~\ref{fig:turbulence}(b) displays the corresponding velocity field \(\bm{u}_{\mathrm{target}}\) for comparison. It is evident that the method effectively recovers the turbulent flow, with the recovered velocity field resembling the target field and capturing key features such as vortex structures and flow direction, demonstrating the algorithm's capability in reconstructing the target velocity field.

\subsection{Three-dimensional knotted vortex filaments}

\begin{figure}
    \centering
    \includegraphics[width=\linewidth]{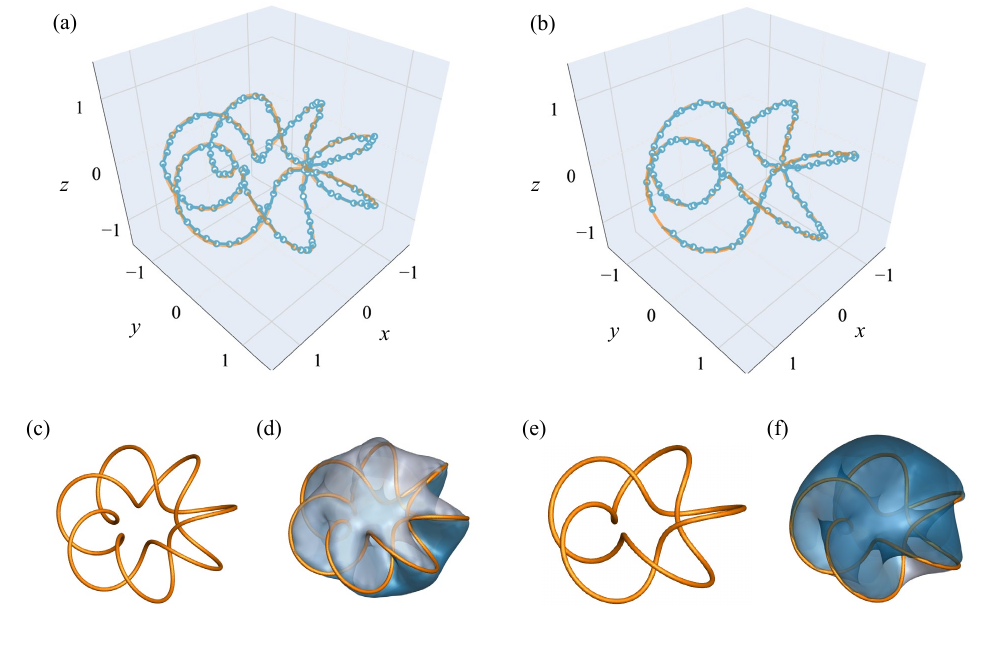}
    \caption{Foil extraction results. (a) (b) The positions of the extracted vortex filaments and target vortex filaments for the septafoil and cinquefoil cases, respectively. The light-blue dots and lines indicates the positions of extracted points and filaments, and the orange line indicates the ground truth. (c) (e) The shapes of the two target vortex filaments. (d) (f) Iso-surfaces of $\mathrm{Im}(\psi) = 0$, colored by the real part of the complex scalar field} $\mathrm{Re}(\psi)$. The zero-intersect lines, representing the vortex filament positions, are marked in orange.
    \label{fig:cinqfoil}
\end{figure}

We further extend our method to three-dimensional knotted vortex filaments, a representative flow characterized by intricate topological structures and rich dynamical behaviour~\citep{Kleckner2013Creation,Barenghi2007Knots,Tao2021}. We construct our knotted vortex filaments according to \citet{xiong2019Construction,Xiong2020}.
The parametric equations of the knot $\boldsymbol{c}(\theta) = \left(c_x (\theta), c_y(\theta), c_z(\theta) \right)$ are
\begin{equation}
    \begin{cases}
        c_x(\theta) = \left(R_t + r_t \cos (q \theta) \right) \cos(p\theta) \\
        c_y(\theta) = \left(R_t + r_t \cos (q \theta) \right) \sin(p\theta) \\
        c_z(\theta) = -r_t \sin(q\theta),
    \end{cases}
    \label{eq:knotvortex}
\end{equation}
where $R_t$ and $r_t$ are the major radii of the knot. $p$ and $q$ represent the shape of the knotted foil. In our case, $R_t = 1$ and $r_t = 0.5$ so the foils are neither too big nor too small that the details would be lost during descritization. We would like to test a septafoil and a cinquefoil, therefore $(p, q)$ is selected to be $(2, 7)$ and $(2, 5)$. The shapes of the two foils are shown in figure~\ref{fig:cinqfoil} (c) and (e) respectively. The computational domain is $(-2,-2, -2) \times (2, 2, 2)$, and is discretized into a $32 \times 32 \times 32$ grid. Referring to the discussion in \ref{sec:sensitivity}, two cases are all optimized with a $N_d = 3500$ circuit depth and 1200 iterations. 

% 介绍一下我们的计算结果。首先我们可以看到(a)(b)两张图体现了目标速度场涡量和提取结果之间的对比，可以看到两者吻合的情况不错。图(d)和图(f)则刻画了波函数虚部为0的等值面，以实部的值着色。可以看到涡丝结果的确沿着波函数的0等值面进行。速度场的MSE分别是0.035和0.057。整体来说表现非常好，我们现在非常有信心（其实没有信心）我们的方法可以用于三位湍流数据的处理与解决。
Figure~\ref{fig:cinqfoil} (a) (b) shows the comparision between target and extracted filament positions, which shows a good match. Figure~\ref{fig:cinqfoil} (d) and (f) shows the isosurface where $\mathrm{Im}(\psi) = 0$, colored by the values of $\mathrm{Re}(\psi)$. It can be seen that the extraction results are located along zero sets of $\psi$. The final MSEs are $0.057$ for septafoils and $0.035$ for cinquefoils , indicating the method’s effective performance in these three-dimensional cases.

\subsection{Three-dimensional turbulent flows}
% 现在我们展示湍流算例。我们的湍流算例从数值计算得出。我们首先在$(-\pi, -\pi, -\pi) \times (\pi, \pi, \pi)$的空间内生成随机的傅里叶相位并以此产生一个初始速度场，然后使用谱方法求解NS方程以进一步保证其各向同性的湍流性质。在此基础上，我们将速度场场离散化为32x32x32的三维网格，并采用电路深度4000并进行了1300步优化进行涡旋提取。对于$h=10$的情况，涡旋提取结果展示在图3DturbA中。可以看到放大图中的涡丝形成了缠绕结构，体现了湍流的性质。
We present results for three-dimensional turbulent flows obtained from numerical simulations. A random initial velocity field is constructed within the domain \([- \pi, \pi]^3\) by assigning Fourier phases to the velocity components. This initial field serves as input to the NS equations, solved using a spectral method to ensure isotropic turbulence properties. The velocity field is then discretized on a uniform $32 \times 32 \times 32$ three-dimensional grid. For vortex extraction, we employ a circuit depth of 4000 and perform 1300 optimization steps. The results for $h = 10$ are shown in figure~\ref{fig:3DTurbA}. The enlarged view reveals entangled vortex filaments, characteristic of turbulent flows.

\begin{figure}
    \centering
    \includegraphics[width=\linewidth]{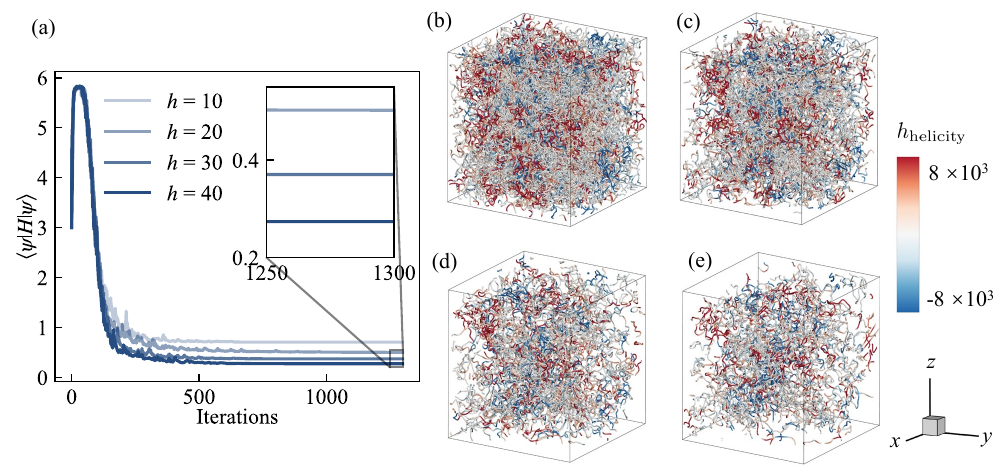}
    \caption{Effect of filament strength \( h \) on convergence and extraction. (a) Convergence of the objective function \(\langle \psi | H | \psi \rangle\), with larger \( h \) yielding smaller final values, indicating improved approximations. (b)--(e) Extracted filaments for \( h = 10 \) (b), \( h = 20 \) (c), \( h = 30 \) (d), and \( h = 40 \) (e). Increasing \( h \) reduces the number of extracted filaments, emphasizing the trade-off discussed in \ref{sec:sensitivity}.} 
    \label{fig:3DTurbB}
\end{figure}

% 我们进一步研究了不同涡强（h值）对结果的影响。如图TurbB (a)所示，尽管四种配置目标函数都趋近于0，但是随着h值的增加，最终收敛值会不断下降。然而，如图TurbB(b)-(e)所示，当h值增大时，涡丝数量会明显减少，表明涡强与涡旋数量之间需要找到合适的平衡点，验证了我们在section中讨论的tradeoff。
The influence of vortex strength, characterized by the parameter $h$, on the results is further analyzed. Figure~\ref{fig:3DTurbB}(a) demonstrates the convergence of the objective function \( \langle \psi | H | \psi \rangle \) over iterations for different filament strengths \( h \). While the objective function for all configurations converges to stable small values, the final convergence value decreases with increasing $h$. As illustrated in figure~\ref{fig:3DTurbB}(b)-(e), the number of vortex filaments diminishes significantly as $h$ increases. This suggests that a balance between vortex strength and filament density must be achieved, corroborating the trade-off discussed in section~\ref{sec:sensitivity}.

\begin{figure}
    \centering
    \includegraphics[width=\linewidth]{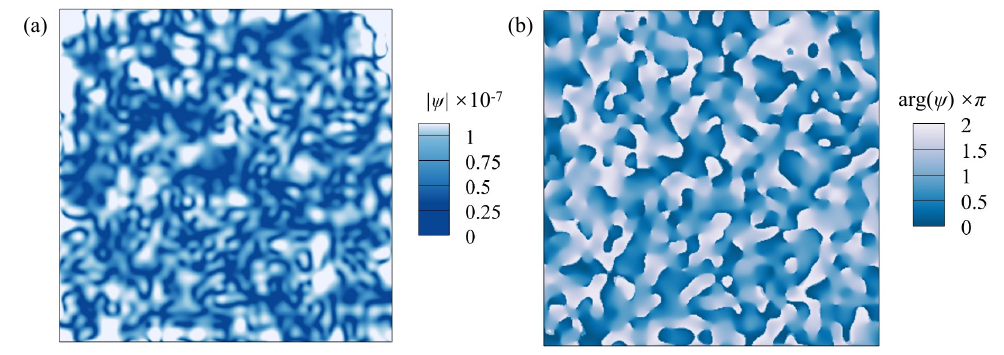}
    \caption{Contours of the scalar field $\psi$ obtained from 3-dimensional turbulences. (a) Contours of the magnitude $|\psi|$ at $y = 0$. (b) Contours of the phase $\arg (\psi)$ at $y = 0$.} 
    \label{fig:3DTurbC}
\end{figure}

% 最后我们展示在几个截面上psi和涡量的分布情况。以h=10为例，我们在x=0, y=0, z=0三个平面上绘制了涡量大小和ψ幅值的contour图。结果表明，ψ的零点对应的涡丝位置与涡量较大的位置大致吻合，进一步验证了ψ与涡量之间的紧密关系。
% Finally, the distributions of $|$ and vorticity $|\omega|$ are examined across selected cross-sections. For $h = 10$, contour plots of $|\psi|$ and $|\omega|$ are shown for the planes $x = 0$, $y = 0$, and $z = 0$. The zero sets of $\psi$ align closely with the positions of the vortex filaments, which correspond to regions of high vorticity. This observation substantiates the strong correlation between $\psi$ and vorticity.

Finally, the distributions of $|\psi|$ and the phase $\arg (\psi)$ are examined across selected cross-sections. For $h = 10$, displayed in \ref{fig:3DTurbC}, contour plots of $|\psi|$ and $\arg (\psi)$ are shown for the planes $y = 0$ and $z = 0$. The results highlight the spatial structure of $\psi$, where the scalar field aligns with underlying coherent features. These observations provide further insights into the relationship between $\psi$ and the flow dynamics.

% 总的来说，本研究提出的方法在三维湍流算例中取得了良好的效果，并为涡旋识别和分析提供了一种新的思路，帮助推进湍流的结构研究。
In summary, the method proposed in this study demonstrates effectiveness in three-dimensional turbulence cases. It provides a novel approach for vortex representation and analysis, contributing to the structural understanding of turbulence.

\section{Conclusions}
\label{sec:conclusion}
This study introduces a quantum implicit representation for the extraction and analysis of vortex filaments in turbulent flows. By reformulating the filament extraction problem as a continuous functional optimization, discretizing it on a staggered grid as a Hermitian eigenvalue problem, solving it with VQE to encode the velocity field in a quantum state, and leveraging a neural network to extract vortex filaments as a reduced-dimensional representation, we offer an efficient approach to overcoming traditional combinatorial challenges, especially in complex flow configurations characterized by high filament densities and intricate topological change. The use of quantum algorithms brings significant computational advantages. Our approach achieves a computational complexity of $\mathcal{O}(N_d \sqrt{N_\mathrm{cell}} \log N_\mathrm{cell})$, which is more efficient than traditional methods such as Gaussian elimination, QR method ($\mathcal{O}(N_\mathrm{cell}^3)$) and the Lanczos algorithm ($\mathcal{O}(m N_\mathrm{cell}^2)$) \citep{Higham2011gaussian, Francis1961qr, Lanczos1950iteration}. The incorporation of machine learning techniques in the post-processing stage enhances the robustness of vortex filament extraction. In the 3D vortex filament scenario, the machine learning offers a 90\% reduce in circuit depth, proving potential in noise reduction and overall efficiency in real-world flows.
% In addition, the incorporation of deep learning techniques in the post-processing stage enhances the robustness of vortex filament extraction, improving noise reduction and overall efficiency in real-world turbulent flows. 
This approach not only advances the understanding of vortex dynamics but also opens up possibilities for its application in areas such as aerodynamics and machine learning-driven fluid dynamics simulations.

Numerical validation through two- and three-dimensional turbulent flows demonstrates the method’s effectiveness. The extracted filaments exhibit good agreement with target configurations, with substantial reductions in mean squared error (MSE) compared to traditional methods. The ability to capture the topological structure of the vortex filaments and align them with coherent flow features highlights the potential of this quantum-based framework as a reliable tool for vortex dynamics research.  

Although this representation method shows significant promise, challenges remain, particularly as we have not yet conducted experiments on quantum hardware, given that quantum computing technology itself is still in its early stages. Future research will focus on further optimizing quantum algorithms, addressing scalability issues, and extending this approach to multi-scale and high-resolution turbulence simulations to adapt to experimental implementation. Additionally, incorporating more advanced machine learning techniques could further enhance the accuracy and efficiency of vortex filament extraction. These advancements will enable quantum methods to more effectively capture the dynamic behavior of vortices in turbulent flows, ultimately improving their applicability in practical applications across engineering and scientific fields.

\section*{Declaration of interests}\label{sec:Dec}
The authors report no conflict of interest.

\section*{Acknowledgments}\label{sec:ack}
The authors acknowledge the support of the National Natural Science Foundation of China (Grants No. 12302294, 12432010 and 12525201) and the National Key Research and Development Program of China (Grant No. 2023YFB4502600). Chenjia Zhu acknowledges the support from the Qizhen Learning Platform of Zhejiang University.

\appendix

\appendix
\section{Pauli operator cutoff}
\label{sec:PauliCutoff}

In Section \ref{sec:VQE_solve}, we discuss how quantum measurements, using Pauli operator decomposition, are employed to optimize quantum circuits and solve eigenvalue problems, with an emphasis on enhancing efficiency through the Pauli operator cutoff method. Named after physicist Wolfgang Pauli, the Pauli operators serve as the fundamental building blocks in quantum mechanics, providing a complete basis for representing Hermitian matrices like Hamiltonians \citep{Landau2013quantum}. The four Pauli operators include the identity matrix \(I\) and the three matrices \(\sigma_x\), \(\sigma_y\), and \(\sigma_z\), which are defined as: 

\begin{equation}
I = \begin{pmatrix}1 & 0 \\ 0 & 1 \end{pmatrix}, \quad 
\sigma_x = \begin{pmatrix}0 & 1 \\ 1 & 0 \end{pmatrix}, \quad 
\sigma_y = \begin{pmatrix}0 & -\mathrm{i} \\ \mathrm{i} & 0 \end{pmatrix}, \quad 
\sigma_z = \begin{pmatrix}1 & 0 \\ 0 & -1 \end{pmatrix}.
\end{equation}

When a Hermitian matrix \(H\) is represented as a linear combination of the tensor products of Pauli operators, it becomes possible to efficiently compute quantum measurements. However, as the number of grid points increases, the complexity of this decomposition grows rapidly. For a grid with \(N_{\text{cell}}\) points, the corresponding Hermitian matrix \(H\) has dimensions \(N_{\text{cell}} \times N_{\text{cell}}\) and can be expressed as a combination of \(N_{\text{cell}}^2\) Pauli operators. This quadratic scaling creates significant computational challenges, especially for large grids.

\begin{figure}
    \centering
\includegraphics[width=\linewidth]{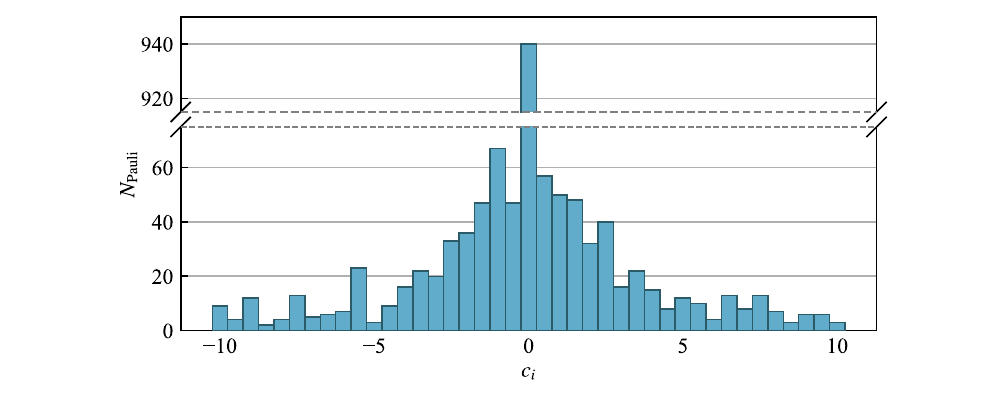}
    \caption {The distribution of coefficients $c_i$ of the decomposition derived from ``Frog leaping'' example. On the horizontal axis, the values of $c_i$ are distributed within a specified range, with each bar spanning a particular interval. The height of each bar corresponds to the number of coefficients that fall within that range. The figure shows that a large proportion of the Pauli operators are contributing small amounts to the objective function, as indicated by the concentration of values around zero.}
    \label{fig:distribution}
\end{figure}

In practical applications, such as analyzing velocity fields, most coefficients in the Pauli decomposition of a Hermitian matrix are negligible. For instance, consider a two-dimensional velocity field with four vortices, illustrated in figure~\ref{fig:winding}, discretized on a \(16 \times 16\) grid. This generates a \(256 \times 256\) Hermitian matrix with 65,536 Pauli coefficients. However, as shown in figure~\ref{fig:distribution} only 1,815 of these coefficients have absolute values greater than \(10^{-5}\). This significant sparsity highlights the potential to simplify the problem by excluding Pauli operators with minimal contributions.

% 柱状图的每个柱子左右两侧会框出一个区间，柱状图的高度代表这个区间内的系数数量。可以看到绝大部分系数都落在0附近，因此很大一部分的Pauli operators都对整体的objective function贡献很少。

The Pauli operator cutoff is a method that takes advantage of the sparsity found in Pauli decomposition. By eliminating operators with negligible contributions, this approach significantly reduces the computational complexity of quantum measurements while maintaining accuracy. The process involves the following steps:

\begin{enumerate}
    \item Decomposition and ranking: The Hermitian matrix \(H\) is represented as shown in \eqref{eq:decompose}. The complex coefficients in this expression are ranked based on their absolute values, ensuring that the most significant terms are prioritized.

    \item Cutoff threshold: A cutoff rate \(r\) is determined to select only the top \(r\)-fraction of the largest coefficients and their associated operators. This step aims to strike a balance between computational efficiency and the required level of accuracy.

    \item Reconstruction: The chosen coefficients and operators are recombined to create a simplified approximation of the original Hermitian matrix, denoted as \(\tilde{H}\). This reduced form retains the essential features of \(H\) while significantly reducing its complexity.
\end{enumerate}

We evaluate the Pauli operator cutoff strategy across various grid sizes and scenarios. The parameter settings for these experiments are adjusted based on grid size: for \(8 \times 8\) grids, \(N_d = 150\) and \(N_i = 60\); for \(16 \times 16\) grids, \(N_d = 300\) and \(N_i = 400\); and for \(32 \times 32\) grids, \(N_d = 300\) and \(N_i = 1000\). These configurations are informed by Section~\ref{sec:sensitivity}.

\begin{figure}
    \centering
\includegraphics[width=\linewidth]{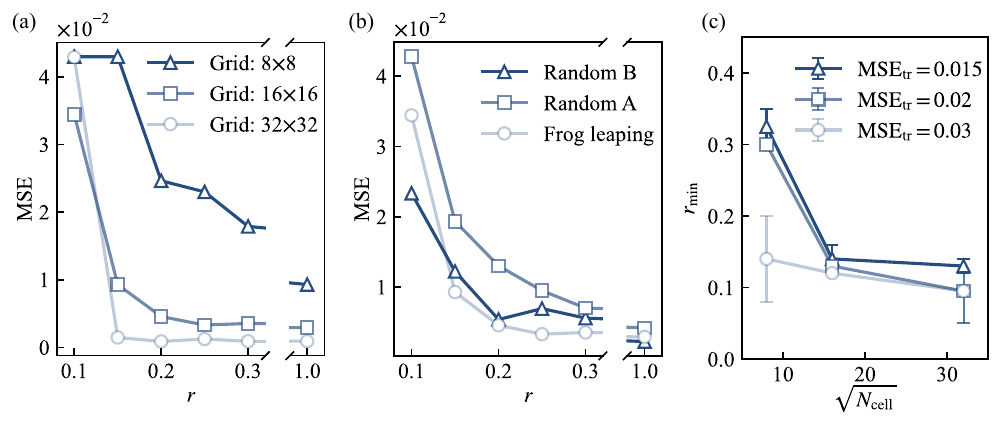}
    \caption{Results of Pauli operator cutoff. (a) MSE against cutoff rate \( r \) across different grid sizes. As the grid gets finer, the ``elbow point'' of cutoff rate decreases. (b) MSE against cutoff rate \(r\) across different cases. Despite the variation of ``elbow point'', all cases see an convergence in MSE as \(r\) increases over 0.3. This indicates the universal validity of the approach.  (c) Minimum retention rate \( r_\mathrm{min} \) required to meet each MSE threshold (\( \mathrm{MSE}_\mathrm{tr} \)) across different grid sizes. The decreasing trend of \( r_\mathrm{min} \) with grid size suggests good scalability of the method.}
    \label{fig:cutoff}
\end{figure}

The results, presented in figure~\ref{fig:cutoff}, demonstrate the effectiveness of the cutoff approach in simplifying computations while maintaining accuracy. In the ``frog leaping'' scenario, shown in figure~\ref{fig:cutoff}(a), we examine grids of sizes \(8 \times 8\), \(16 \times 16\), and \(32 \times 32\), and analyze the mean squared error (MSE) between the reconstructed and ground-truth velocity fields as a function of the cutoff rate \(r\). For the \(8 \times 8\) grid, a cutoff rate retaining approximately 30\% of the Pauli operators achieves satisfactory accuracy. For larger grids (\(16 \times 16\) and \(32 \times 32\)), a lower cutoff rate of around 15–20\% is sufficient to attain similar performance. Other scenarios are evaluated using a \(16 \times 16\) grid, as shown in figure~\ref{fig:cutoff}(b). This further illustrates the variation in the ``elbow point,'' where the MSE stabilizes across different scenarios, highlighting the adaptability of the cutoff method to diverse situations. We further investigated the minimum selectable rate \( r_{\min} \) as a function of \( \sqrt{N_\mathrm{cell}} \) under different MSE thresholds \( \mathrm{MSE}_\mathrm{tr} \). For each grid resolution, experiments were conducted on both the ``Frog leaping'' and ``Random example A'' cases, with \( \mathrm{MSE}_\mathrm{tr} \) set to 0.015, 0.02, and 0.03. The average \( r_{\min} \) across the two cases is used to represent the minimum rate required for each grid--threshold pair. The results, shown in Figure~\ref{fig:cutoff}(c), plot the averages as data points, with error bars indicating the variation between cases. A clear downward trend is observed as the grid resolution increases. Notably, on the \(32 \times 32\) grid, \(r_{\min}\) falls below 0.1 for certain values of \(\sqrt{N_{\text{cell}}}\) when \(\mathrm{MSE}_{\mathrm{tr}} = 0.03\), demonstrating the scalability of the proposed method. Overall, the results suggest that the Pauli operator cutoff strategy is a promising tool for reducing computational complexity without compromising accuracy.

The Pauli operator cutoff offers a practical approach to reducing the computational cost of quantum measurements by leveraging the sparsity in Pauli decompositions. It achieves significant efficiency gains while maintaining accuracy, making it a valuable tool for scaling quantum algorithms. However, the selection of an optiamal cutoff rate $r$ remains problem-dependent, as different scenarios and grid setups derive different Hermitian matrix $H$. Future efforts could explore adaptive cutoff techniques to further refine the balance between efficiency and precision across various applications.

\section{A data-driven approach to vortex filament extraction}
\label{sec:NNextraction}
\begin{figure}
    \centering
    \includegraphics[width=\linewidth]{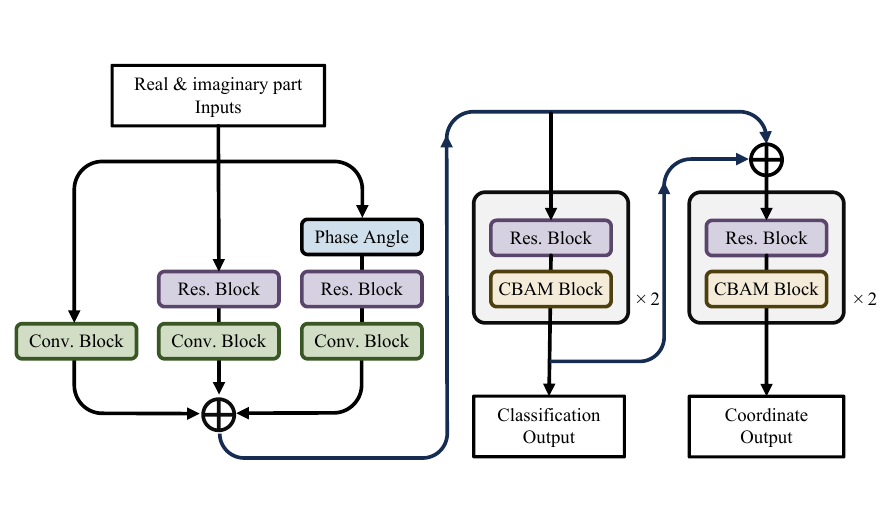}
    \caption{Illustration of the proposed neural network architecture for vortex filament extraction. The network consists of three primary branches to process the scalar field input, which are combined to produce an initial feature representation. Classification outputs, predicting the presence and winding numbers of vortex filaments, are generated in parallel with coordinate inference outputs, which estimate the local start and end points of vortex filaments in each grid cell. Residual blocks and CBAM (Convolutional Block Attention Module) modules \citep{woo2018cbam} are integrated to enhance feature learning and improve prediction accuracy for noisy input data. “Conv. block” denotes a convolutional block.}
    \label{fig:network}
\end{figure}
\subsection{General framework}
Section~\ref{sec:extract} provides a classical method for extracting the zero set of scalar field, which in our case is the vortex filament. However, this method is limited when handling coarse grids or data influenced by errors from VQEs. To overcome these limitations, we designed a deep learning-based method to efficiently extract the zero set of the scalar field from noisy data.

Our network processes the scalar field $\psi$ in three-dimensional space, aiming to detect whether there are vortex filaments crossing a hexahedral grid and, if so, to predict the start and end coordinates of the vortex filaments. The network input consists of discretized scalar field data with a shape of $(B, C_{\text{in}}, N_x, N_y, N_z)$, where $B$ is the batch size, $C_{\text{in}}=2$ denotes the number of input channels (representing real and imaginary components of the complex field), and $N_x, N_y, N_z$ are the grid resolutions along the three spatial dimensions. 
% The network input consists of discretized scalar field data with a shape of $(B, 2, N_x, N_y, N_z)$, where $B$ is the batch size and $N_x, N_y, N_z$ are the grid resolutions along the three dimensions. 

The network output contains two components with the following tensor structures. Classification output predicts winding numbers and vortex existence across grid faces. The output shape is \( (B, C_{\text{nw}}, F, N_x - 1, N_y - 1, N_z - 1) \), where \( C_{\text{nw}} = 3 \) corresponds to three winding number classes (\( n_w = -1, 0, +1 \)); \( F = 6 \) represents the six faces of each hexahedral cell; and the spatial dimensions \( (N_x - 1) \times (N_y - 1) \times (N_z - 1) \) index the grid cells. Coordinate inference output predicts normalized local coordinates of vortex endpoints. The output shape is \( (B, D, S, N_x - 1, N_y - 1, N_z - 1) \), where \( D = 3 \) specifies three-dimensional spatial coordinates (x, y, z), and \( S = 2 \) denotes the start and end points of vortex segments.

% Classification output (winding number and vortex existence): Indicates whether the six faces of each hexahedral grid cell are crossed by vortex filaments, corresponding to $n_w = -1, 0, +1$ for three classes. The output shape is $(B, 3, 6, N_x-1, N_y-1, N_z-1)$, where $n_w \neq 0$ indicates that the grid cell contains a vortex filament.

%Coordinate inference output: Predicts normalized local coordinates of vortex endpoints. The output shape is $(B, D, S, N_x-1, N_y-1, N_z-1)$ where: $D=3$ specifies three-dimensional spatial coordinates (x, y, z); $S=2$ denotes start and end points of vortex segments.

% Coordinate inference output: Predicts the normalized local coordinates of the start and end points of the vortex filament in the grid cell where crossing occurs. The output shape is $(B, 3, 2, N_x-1, N_y-1, N_z-1)$.

For the classification output, a grid cell is considered to contain a vortex filament if any of its six faces is classified as “crossed by a vortex filament”($n_w \neq 0$). The coordinate prediction branch only produces physically meaningful results when the corresponding grid cell contains vortex filaments. Complete vortex filaments can be reconstructed by connecting predicted segments from adjacent grid cells.
% The position prediction branch only has physical significance when the classification branch identifies the grid cell as containing a vortex filament. To reconstruct complete vortex filaments, these vortex segments can be connected by considering predictions from adjacent grid cells.

The network architecture is illustrated in figure~\ref{fig:network}. The input data undergoes parallel feature extraction through three independent branches, which are subsequently fused into a unified feature representation. This initial feature tensor is processed by the classification module to generate winding number predictions. The classification output then diverges into two streams: one directly serves as the final classification result, while the other combines with the initial features through skip connections before entering the coordinate regression module for endpoint prediction.

%因为结论里要有我们方法的定量叙述，所以我还是要把这个方法的训练数据、vqa的参数写在这里。这些参数其实是和正文的参数有冲突的，因此之前被略去不表。希望不会有人来看我们的附录。
The training and testing datasets consist of random knotted filaments generated based on the equation \eqref{eq:knotvortex}. The parameters are randomly selected to produce different knot shapes. Each case is discretized into a $16 \times 16 \times 16$ grid, and the velocity field is computed using the Biot-Savart law. For each case, the circuit depth is set to $N_d = 200$, and optimization is performed for $N_i = 2000$ iterations. Compared to an optimal $N_d = 2000$ discussed in \ref{sec:sensitivity}, the circuit depth in the machine learning datasets is intentionally reduced to simulate scenarios with noise and error. It's also common to reduce circuit depth for noise control in quantum computing \citep{Bharadwaj2024Simulating}.

\subsection{Loss function}
Our neural network is trained using a composite loss function comprising five components (the winding number classification loss $\mathcal{L}_{w_n}$, the vortex existence classification loss $\mathcal{L}_{\text{exist}}$, the coordinate inference loss $\mathcal{L}_{\text{coord}}$, the isolated grid flux conservation loss $\mathcal{L}_{\text{flux}}$, and the adjacent grid flux conservation loss $\mathcal{L}_{\text{symmetry}}$) that enforce both data fidelity and physical consistency. The total loss is formulated as:
\begin{equation}
\mathcal{L}_{\text{total}} = \lambda_1 \mathcal{L}_{\text{$w_n$}} + \lambda_2 \mathcal{L}_{\text{exist}} + \lambda_3 \mathcal{L}_{\text{coord}} + \lambda_4 \mathcal{L}_{\text{flux}} + \lambda_5 \mathcal{L}_{\text{sym}}
\end{equation}
where $\{\lambda_i\}_{i=1}^5$ are tunable hyperparameters controlling the relative importance of each objective. In our training strategy, we adopt a phased approach with dynamic weighting parameters to address the challenges posed by the coupled objectives in our loss function. Initially, we set $\lambda_2$ to 0 while assigning a value of 1 to the other weights. This configuration ensures that the network focuses exclusively on learning stable winding number features under the guidance of physical constraints, namely, the isolated grid flux conservation and adjacent grid flux conservation losses. Based on this training approach, the model establishes reliable per-face predictions for the winding numbers, which indirectly inform the vortex coordinate predictions.
In fact, our ablation studies indicate that without this early focus on winding number features, the model bypasses the learning of the essential characteristics of the faces and directly focuses on the vortex existence information determined by these features. This behavior leads to an over-reliance on coarse, high-level existence signals while neglecting the fine-grained local details crucial for precise coordinate regression, thus resulting in suboptimal performance and diminished robustness.
Once the loss begins to converge, signaling that the winding number features have matured sufficiently to support robust existence determination, we reintroduce the vortex existence classification by setting $\lambda_2 = 5$. Concurrently, we adjust $\lambda_3$ to 15 to further enhance the accuracy of coordinate regression. This deliberate dynamic reweighting strategy not only amplifies the gradient signals for both the existence classification and coordinate regression tasks but also balances the competing objectives in the extraction process.
 
\begin{figure}
    \centering
    \includegraphics[width=10cm]{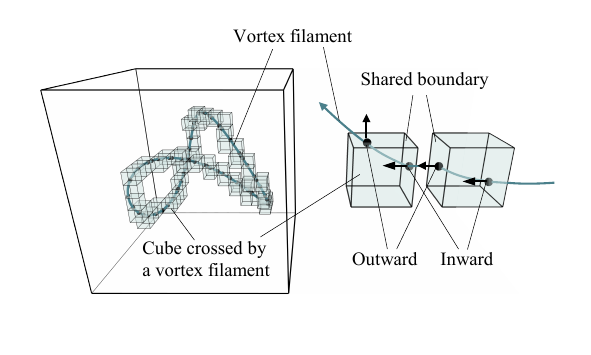}
    \caption{Illustration of vortex filaments passing through grid cells. The left diagram shows vortex filaments intersecting multiple grid cells, where each cube is crossed by one or more vortex filaments. The right diagram highlights the shared boundaries between adjacent cubes, with arrows indicating the inward and outward flux. These visualizations support the implementation of flux conservation and symmetry constraints in the loss functions $\mathcal{L}_{\text{flux}}$ and $\mathcal{L}_{\text{symmetry}}$, ensuring physical consistency of the predicted vortex filaments.}
    \label{fig:cube and filament}
\end{figure}

(1) Classification losses for winding number and vortex existence: These two classification tasks exhibit intrinsic coupling but divergent convergence behaviors, necessitating differentiated treatment. Let $\mathcal{L}_{w_n}$ denote the winding number classification loss for grid faces, and $\mathcal{L}_{\text{exist}}$ represent the vortex existence loss for grid cells. Their relationship follows from topological constraints: a cell contains vortices if and only if at least one face exhibits non-zero winding number.

To handle severe class imbalance (most faces have $n_w=0$ and cells contains no vortex filament), both losses employ the Focal loss \citep{ross2017focal} to stabilize training and avoid local optima. For each classification task, we employ a basic Focal loss defined as:  
\begin{equation}
\mathcal{L}_{\text{Focal}} = -\alpha_t (1 - p_t)^\gamma \log(p_t),
\end{equation}
where $p_t$ is the predicted probability of the true class, $\alpha_t$ is a balancing factor, and $\gamma$ is a focusing parameter that reduces the relative loss contribution from well-classified examples.

Despite the shared intrinsic geometric relationship, empirical observations reveal differing convergence dynamics between the tasks, which is one of the reasons why we need to adopt a phased dynamic weighting strategy for the parameters.
% \sout{To address the coupled objectives, we implement a phased training strategy. In the initial phase, we disable the loss term \(\mathcal{L}_{\text{exist}}\) (i.e., set \(\lambda_2 = 0\)) for several epochs before attempting to refine the existence classification. This allows the model to establish reliable per-face predictions for the winding numbers, which indirectly inform the vortex coordinates. Once the model begins to converge, we reactivate \(\mathcal{L}_{\text{exist}}\) by setting \(\lambda_2 = \lambda_1\) and increasing \(\lambda_3\) (as described below) to emphasize vortex existence accuracy and coordinate inference. This staged approach prevents the existence classifier from overfitting to premature face predictions while ensuring that the winding number features have matured sufficiently to support robust existence determination.}

(2) Coordinate regression: 
For the regression task of inferring vortex coordinates, we use an MSE loss computed only for cells that are determined to contain vortex filaments (based on the existence prediction). Each such cell provides two coordinate targets: one for the starting point and one for the ending point of the vortex segment. The overall coordinate loss is defined as:
\begin{equation}
\mathcal{L}_{\text{coord}} = \mathbb{E}_{\substack{\text{cells with vortex}}} \left[ \left(y^{(s)} - \hat{y}^{(s)}\right)^2 + \left(y^{(e)} - \hat{y}^{(e)}\right)^2 \right],
\end{equation}
where the expectation $\mathbb{E}_{\substack{\text{cells with vortex}}}[\cdot]$ is taken only over the cells that contain vortex filaments.
Here, $y^{(s)}$ and $ y^{(e)} $denote the ground-truth coordinates for the start and end points, and $\hat{y}^{(s)}$ and $\hat{y}^{(e)}$ are the corresponding predicted coordinates.

(3) Physical constraints (Flux conservation and symmetry constraints): 
While the aforementioned losses focus on classification and localization, ensuring the physical consistency of vortex filaments is equally crucial. To address this, we introduce two additional loss functions, $\mathcal{L}_{\text{flux}}$ and $\mathcal{L}_{\text{sym}}$, which do not rely on labeled ground-truth data but instead impose physical constraints on the predicted fields. By enforcing these structural regularities, these losses ensure that the resulting vortex filaments remain continuous and adhere to flux conservation principles.

% 上一个版本的表达式：
% 1. Flux conservation loss $\mathcal{L}_{\text{flux}}$:  
%    Each cube in the spatial grid is bounded by six faces (see figure~\ref{fig:cube and filament}). Conceptually, for a physically valid vortex configuration, the net vortex flux entering a cube should equal the net flux exiting it. To enforce this, we define:
%    \begin{equation}
%    \mathcal{L}_{\text{flux}} = \left( \sum_{i=1}^{6} \sum_{j=1}^{3} \beta_j p_{ij}\right)^2,
%    \end{equation}
%    where $i \in \{1,\ldots,6\}$ indexes the cube’s faces (e.g., left, front, bottom, right, back, top), $j \in \{1,2,3\}$ indexes the predicted classes corresponding to $w_n = -1, 0, 1$, $\beta_j$ is the coefficient associated with each winding number value, and $p_{ij}$ is the predicted probability that face $i$ belongs to class $j$. Minimizing this term encourages the total inward and outward flux to balance out, reflecting proper conservation.
   
Each cell in the spatial grid is bounded by six faces (see figure~\ref{fig:cube and filament}), and the net flux through its six faces must vanish:
\begin{equation}
\mathcal{L}_{\text{flux}} = \mathbb{E}_{\text{cells}} \left[\left(\sum_{f=1}^6 \sum_{c=-1}^1 \nu_c p_{f,c}\right)^2\right]
\end{equation}
where $\mathbb{E}_{\text{cells}}[\cdot] $ represents the average of the enclosed expression computed over all cells, $\bm{\nu} = [-1, 0, 1]$ encodes the flux contribution per winding number class $c$, and $p_{f,c}$ is the predicted probability for face $f$ belonging to class $c$.

% 上一个版本的表达式：
% 2. Symmetry constraint loss $\mathcal{L}_{\text{symmetry}}$:  
%    Similarly, the flux crossing a shared boundary between two adjacent cubes should be symmetric, as illustrated in figure~\ref{fig:cube and filament}. For each face $i$, there is a corresponding opposite face $k$ in the neighboring cube. Likewise, there is a symmetry in the winding numbers, where the classes $j=1,2,3$ (for $w_n=-1,0,1$) map to corresponding classes $l=3,2,1$ on the adjacent face. To ensure this balance, we define:
%    \begin{equation}
%    \mathcal{L}_{\text{sym}} = \sum_{i=1}^{6}\left(\sum_{j=1}^{3} (\beta_j p_{ij} + \beta_l p_{kl})\right)^2,
%    \end{equation}
%    where $k$ and $l$ are the indices of the opposite face and the mapped class, respectively. Minimizing this term encourages symmetric flux distributions across shared interfaces, further reflecting the physical continuity and consistency of vortex filaments.
   
Similarly, the flux crossing a shared boundary between two adjacent cubes should be symmetric, as illustrated in figure~\ref{fig:cube and filament}. Adjacent cells sharing face $f$ must agree on the interface flux:
\begin{equation}
\mathcal{L}_{\text{sym}} = \mathbb{E}_{\text{faces}} \left[\left(\sum_{c=-1}^1 \nu_c p_{f,c} + \sum_{c=-1}^1 \nu_{-c} q_{\bar{f},c}\right)^2\right]
\end{equation}
where $\mathbb{E}_{\text{faces}}[\cdot] $represents the average computed over all faces and $q_{\bar{f},c}$ denotes the probability distribution on the opposing face $\bar{f}$ of the neighboring cell.

% \subsection{Dataset and results}
% We randomly generated a series of 3D vortex filament cases and computed the corresponding velocity fields based on the Biot-Savart law, storing the results in a $16 \times 16 \times 16$ grid. Furthermore, we  simulated a noisy scenario characterized by "shallow depth" and "incomplete convergence." The wave function data obtained from measurements served as the input to our network.

% For a $16^3$ grid, achieving good convergence theoretically requires a complexity of \zhu{$\dots$}, equivalent to iterating the 200-layer circuit \zhu{$\dots$} times. Our noise simulation effectively reduced the number of iterations by $\dots$. Under such conditions, directly applying classical methods to extract vortex filaments from the wave function yields results as shown in Figure~\ref{fig:NN and Cla}(a). The extracted vortex filaments exhibit an overall deviation from the ground truth.

% In contrast, utilizing our proposed network method allows for an almost perfect reconstruction of the true vortex filaments. Moreover, the computational complexity of the network inference scales linearly with the number of grid points, consistent with that of classical methods, without a significant increase in computational overhead.

\bibliographystyle{jfm}
\bibliography{main}

\end{document}